\documentclass[12pt]{scrartcl}
\usepackage[margin=2.5cm]{geometry}

\usepackage{array}
\usepackage{tabularx}
\usepackage[affil-it]{authblk}
\usepackage{moreverb}
\usepackage{hyperref}
\usepackage{amsmath}
\usepackage{amsthm}
\usepackage{amssymb}
\usepackage{bm}
\usepackage{color}
\usepackage{graphicx}
\usepackage{enumerate}
\usepackage{microtype}
\usepackage{nicefrac}
\usepackage{booktabs}
\usepackage[T1]{fontenc}
\usepackage[font=small,labelfont=bf,tableposition=top]{caption}
\usepackage[toc,page]{appendix}
\usepackage{algorithm}
\usepackage{algorithmic}
\usepackage[affil-it]{authblk}
\usepackage[numbers]{natbib}

\begin{document}

\providecommand{\keywords}[1]{\textbf{\textit{Keywords: }} #1}
\title{Stochastic Modelling of Urban Structure\footnote{Ellam L, Girolami M, Pavliotis GA,Wilson A. 2018 Stochastic modelling of urban structure. Proc. R. Soc. A 20170700.  \url{http://dx.doi.org/10.1098/rspa.2017.0700}}}
\date{2 October 2017}
\author[1, 2]{L. Ellam\thanks{Email: \href{mailto:l.ellam16@imperial.ac.uk}{l.ellam16@imperial.ac.uk}}}
\author[1, 2]{M. Girolami}
\author[1]{G. A. Pavliotis}
\author[2, 3]{A. Wilson}
\affil[1]{Department of Mathematics, Imperial College London, London, SW7 2AZ}
\affil[2]{The Alan Turing Institute, The British Library, London, NW1 2DB}
\affil[3]{Centre for Advanced Spatial Analysis, University College London,  London, W1T 4TJ}

\maketitle

\begin{abstract}
The building of mathematical and computer models of cities has a long history. The core elements are models of flows (spatial interaction) and the dynamics of structural evolution. In this article, we develop a stochastic model of urban structure to formally account for uncertainty arising from less predictable events.  Standard practice has been to calibrate the spatial interaction models independently and to explore the dynamics through simulation.  We present two significant results that will be transformative for both elements. First, we represent the structural variables through a single potential function and develop stochastic differential equations (SDEs) to model the evolution. Secondly, we show that the parameters of the spatial interaction model can be estimated from the structure alone, independently of flow data, using the Bayesian inferential framework.  The posterior distribution is doubly-intractable and poses significant computational challenges that we overcome using Markov chain Monte Carlo (MCMC) methods. We demonstrate our methodology with a case study on the London retail system.
\end{abstract}


\section{Introduction}
\label{S:int}
The task of understanding the inner workings of cities and regions is a major challenge for contemporary science.  The key features of cities and regions are activities at locations, flows between locations and the structure that facilitates these activities~\cite{dearden2015explorations}.  It is well understood that cities and regions are complex systems, and that an emergent structure arises from the actions of many interacting individuals.  The flows between locations arise from the choices of individuals.  An understanding of the underlying choice mechanism is therefore advantageous for planning and decision making.  Economists have long supported the idea that consumer choices are derived from utility, a measure of net benefit, although preferences can only be measured indirectly by the phenomena they give rise to~\cite{marshall1920principles}.

Random utility models, such as the multinomial logit model~\cite{mcfadden1973conditional}, provide a discrete choice mechanism based on a utility function.  These models have received considerable attention in the econometrics literature~\cite{baltas2001random}.   The more conventional random utility models assume that choices are conditionally independent and require large volumes of flow data to calibrate.  It is generally difficult to ascertain the flow data for a large number of individuals residing in a country or city, and this may require an extensive survey that suffers from sampling biases.  On the other hand, the structure facilitating activities can be more straight-forward to measure.

It turns out that the flows between locations concern a vast number of individuals and are well-represented by statistical averaging procedures~\cite{wilson1967statistical}.  It also turns out that the evolution of urban structure can be described by a system of coupled first-order ordinary differential equations that are related to the competitive Lotka-Volterra models in ecology~\cite{harris1978equilibrium}.  The conventional Harris and Wilson model in~\cite{harris1978equilibrium} is obtained by combining Lotka-Volterra models with statistical averaging procedures, after having expressed the flows in terms of the evolving structure and spatial interaction.  As it tends to be more feasible to observe the emergent structure, for example, configurations of floorspace  dedicated to retail activity, our work is largely motivated by the existing models of urban structure~\cite{dearden2015explorations, harris1978equilibrium, wilson2000complex, wilson2011entropy, wilson2012science}.  By adopting a similar approach, we view the flows between locations as `missing data'.

We note, however, that there is an urgent need to provide an improved modelling capability that captures the stochastic nature and uncertainty associated with the evolution of urban structure.  The key shortcoming of the Harris and Wilson model is that it is deterministic and converges to one of multiple equilibria as determined by the initial conditions.  In reality, the behaviour provided by the Harris and Wilson model would be accompanied by fluctuations arising from less predictable events.  We instead introduce a mathematically well-posed systems of stochastic differential equations (SDEs) to address this shortcoming, and provide an associated Bayesian inference methodology for parameter estimation and model calibration.

To this end, we take a novel approach and construct a probability distribution to represent the uncertainty in equilibrium structures for urban and regional systems.  The probability distribution is a Boltzmann-Gibbs measure that is the invariant distribution of a related SDE model~\cite{pavliotis2016stochastic}, and is defined in terms of a potential function whose gradient describes how we expect urban structure to evolve forward in time.  The potential function may be interpreted as constraints on consumer welfare and running costs from a maximum entropy argument~\cite{wilson2011entropy, lasota2013chaos}.  For the purposes of parameter estimation, the Boltzmann-Gibbs measure forms an integral part of the assumed data generating process in a Bayesian model of urban structure~\cite{kaipio2006statistical, gelman2014bayesian}. A computational statistical challenge arises as there is an intractable term in the density of the Boltzmann-Gibbs measure that is parameter-dependent.  The intractable term must be taken into consideration when using Markov chain Monte Carlo (MCMC) to explore the probability distributions of interest~\cite{liu2008monte, murray2006}.  Our approach is applicable to a wide range of applications in urban and regional modelling;  we demonstrate our approach by inferring the full distribution over the model parameters and latent structure for the London retail system.

	\section{Modelling urban systems}
\label{S:p_meas_urban_systems}

In this section, we construct a probability distribution for urban and regional systems.  We work in the setting of the Harris and Wilson model~\cite{harris1978equilibrium} and use consumer behaviour as an archetype, however, the methodology is general and has wider applications such as archaeology, logistics, health care and crime to name a few~\cite{dearden2015explorations}.  We are interested in the sizes of $M$ destination zones where consumer-led activities take place, for example, shopping.  Similarly, there are $N$ origin zones from where consumers create demands for each of the destination zones.  We define urban structure as the vector of sizes $W = \{ W_1, \dots, W_M \} \in \mathbb{R}_{>0}^M$.  In what proceeds it is more natural to work in terms of log-sizes $X = \{ X_1, \dots, X_M \} \in \mathbb{R}^M$ where each $W_j = \exp(X_j)$.  We refer to log-size as the attractiveness, which is an unscaled measure of benefit, and by working in terms of attractiveness we avoid positivity issues when developing a stochastic model.  We first describe a stochastic generalisation of the Harris and Wilson model and then consider the equilibrium distribution as a probability distribution of urban structure.

\subsection{A stochastic reformulation of the Harris and Wilson model}

\begin{figure}
	\centering
	\includegraphics[scale=0.4]{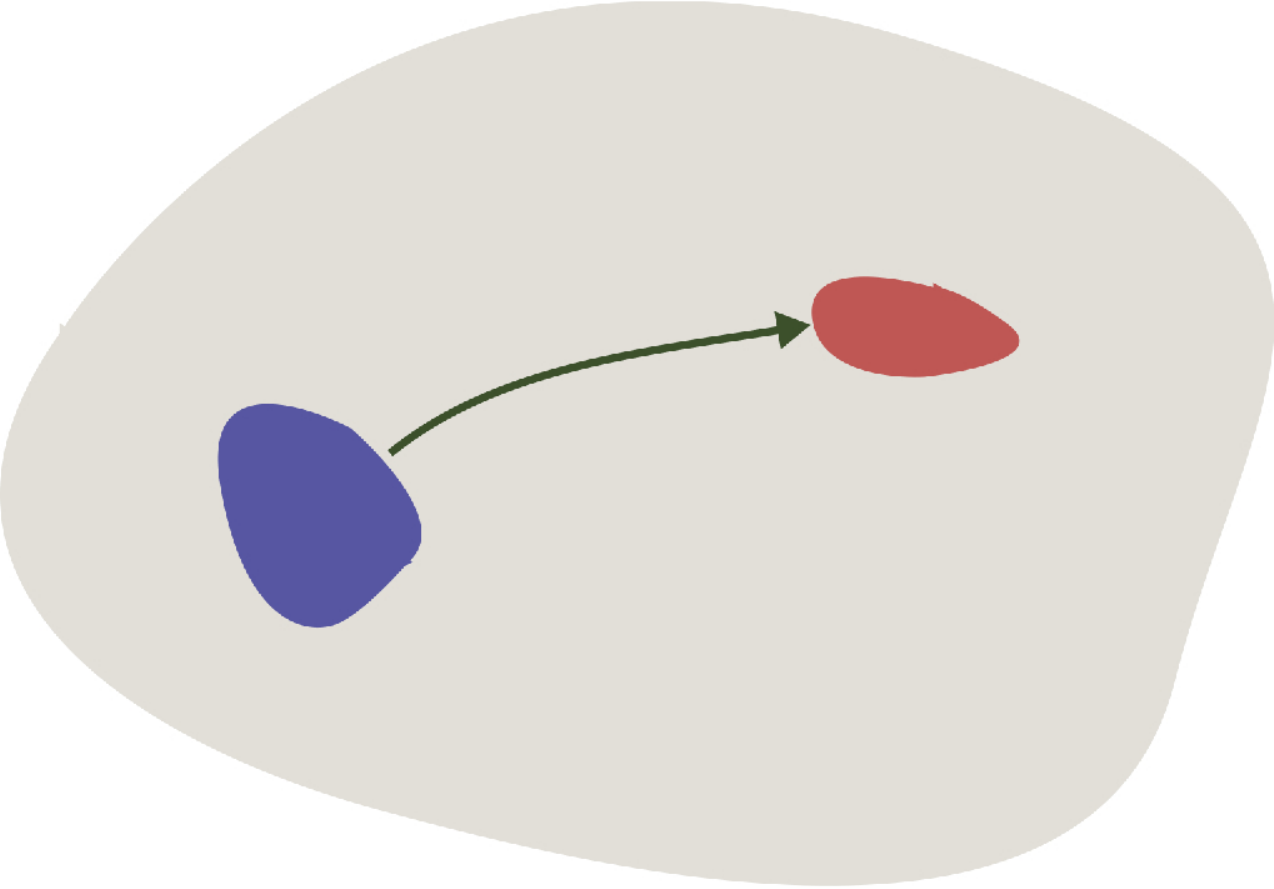}
	\put(-85,67){$T_{ij}$}
	\put(-110,20){$O_{i}$}
	\put(-40,50){$D_{j}$}
	\caption{Illustration of a flow in a urban or regional system.  It is assumed that there are $N$ origin zones (for example, left) and $M$ destination zones (for example, right).  The flow $T_{ij}$ denotes the flow of quantities from origin zone $i$ to destination zone $j$.  In a urban or regional system there are $NM$ flows similar to the one depicted.}
	\label{fig:system}
\end{figure}

The flow of between destination zone $j$ and origin zone $i$ is denoted $T_{ij}$.  We illustrate a component of an urban or regional system in Figure~\ref{fig:system}.  For a singly-constrained urban system, the demands made by the $N$ origin zones are
\begin{equation}
\label{o_eq}
O_i = \sum_{j=1}^M T_{ij}, \quad i = 1, \dots, N,
\end{equation}
and are known. The demands made for the $M$ destination zones are
\begin{equation}
\label{d_eq}
D_j = \sum_{i=1}^N T_{ij}, \quad j = 1, \dots, M,
\end{equation}
and are to be determined.  The demands for the destination zones depend on urban or regional structure.  It is assumed that larger zones provide more benefits for their use and that local zones are more convenient and cost less to use.  A suitable model of the flows is obtained by maximizing an entropy function subject to the constraint in~\eqref{o_eq} in additional to fixed benefit and cost constraints~\cite{wilson1967statistical, wilson2011entropy}.  The resulting flows are
\begin{equation}
\label{flow}
T_{ij} = O_i \frac{  W_j^\alpha \exp (-\beta c_{ij})}{\sum_{k=1}^M W_k^\alpha\exp(- \beta c_{ik})},
\end{equation}
where $\alpha, \beta > 0$ are scaling parameters and each $c_{ij} \geq 0$ represents the cost or inconvenience of carrying out an activity at zone $j$ from $i$.

We expect that zones with unfulfilled demand will grow, whereas zones that do not fulfil their capacity will reduce to a more sustainable size.  It is therefore reasonable to expect a degree of stability in the sizes of the destination zones.  A suitable model of the dynamics is given by the Harris and Wilson model~\cite{harris1978equilibrium}, which is described by a system of ordinary differential equations (ODEs)\footnote{For simplicity, we are not explicit about time-dependence when describing differential equations. }
\begin{equation}
\label{HW}
\frac{dW_j}{dt} = \epsilon W_j \big(D_j - \kappa W_j \big), \qquad W(0) = w_0,
\end{equation}
where $\epsilon > 0$ is the responsiveness parameter and $\kappa>0$ is the cost per unit floor size.  The assumption that  zones aim to maximize their size until an equilibrium is reached is justified by including the cost of capital in the running costs.  A natural generalisation of the Harris and Wilson model is the following SDE with multiplicative noise that we interpret in the Stratonovich\footnote{The Stratonovich interpretation is obtained from a smooth approximation of white noise, which is appropriate for the dynamical system of interest.  For further discussion refer to~\cite{pavliotis2016stochastic}.} sense
\begin{equation}
\label{shw}
{dW_j} = \epsilon W_j \big(D_j - \kappa W_j \big) dt + \sigma W_j \circ {dB_j}, \qquad W(0) = w_0,
\end{equation}
for a standard $M$-dimensional Brownian motion $B$ and volatility parameter $\sigma > 0$.  A heuristic interpretation of the SDE is that over a short time $\delta t$, the net capacity term ${``D_j-\kappa W_j"}$ in~\eqref{HW} is randomly perturbed by centred Gaussian noise with a standard deviation of $\sigma\sqrt{\delta t}$.  The noise term represents fluctuations in the growth rates arising from less predictable events that are not captured by the original model.  The specification of multiplicative noise preserves the positivity of each $W_j$.

With the change of variables $X_j = \ln W_j$, the Harris and Wilson model in~\eqref{HW} can be expressed as a gradient flow.  The corresponding stochastic dynamics in~\eqref{shw} is an overdamped Langevin diffusion.  To express this notion, we introduce a potential function $V:\mathbb{R}^M \to \mathbb{R}$, its gradient $\nabla V : \mathbb{R}^M \to \mathbb{R}^M$ and an `inverse-temperature' parameter $\gamma = 2\sigma^{-2}$ and reformulate the stochastic dynamics as
\begin{equation}
\label{lang}
{dX} = -\nabla V(X) dt +  \sqrt{2 \gamma^{-1}} {dB}, \qquad X(0) = x_0,
\end{equation}
where the potential function is
\begin{equation}
\label{hw_pot}
\epsilon^{-1} V(x) = -\alpha^{-1} \sum_{i=1}^N {O_i}  \ln \sum_{j=1}^M \exp \big( \alpha x_j - \beta c_{ij}) \big) + \kappa \sum_{j=1}^M \exp(x_j).
\end{equation}
It is well understood that the density of $X(t)$, denoted $\rho(x, t)$, evolves in time according to the Fokker-Planck equation~\cite{pavliotis2016stochastic}.  For the SDE in~\eqref{lang}, the Fokker-Planck equation can be written as
\begin{equation}
\label{FP}
\frac{d\rho(x, t)}{dt} = \nabla \cdot \big(\rho(x, t) \nabla V(x) \big) +  \gamma^{-1}\Delta \rho(x, t), \qquad \rho(x, 0) = \delta(x - x_0).
\end{equation}
Whilst~\eqref{FP} is very challenging to solve, especially in higher-dimensions, its steady-state solution is available in closed form and is the density of a Boltzmann-Gibbs measure given by
\begin{equation}
\label{BoltzmannGibbs}
\rho_\infty(x) = \frac{1}{Z} \exp\big(-\gamma V(x) \big), \qquad Z := \int_{\mathbb{R}^M} \exp\big(-\gamma V(x) \big) dx.
\end{equation}
The Boltzmann-Gibbs measure described by~\eqref{BoltzmannGibbs} forms the basis of our stochastic model of urban structure.  The potential function given by~\eqref{hw_pot} does not yield a well-defined probability distribution as the normalizing constant in~\eqref{BoltzmannGibbs} is not finite.  In order to address the issue we could restrict the dynamics to a bounded subset of $\mathbb{R}^M$, or introduce a confining term in the potential function.  We adopt the latter approach and later argue that this approach amounts to an economically meaningful constraint.

\subsection{Boltzmann-Gibbs measures for urban structure}
We model urban and regional structure as a single realisation of the Boltzmann-Gibbs measure described by~\eqref{BoltzmannGibbs}.  The Boltzmann-Gibbs measure is the stationary distribution of the overdamped Langevin dynamics considered, however, we acknowledge that there are other stochastic processes that have the same stationary distribution~\cite{duncan2016variance}.  It is desirable that the potential function satisfies the assumptions in Appendix~\ref{Appendix:Assumptions}.  It suffices to say here that smooth potential functions that grow at least linearly but no faster than exponentially at infinity have the desired mathematical properties.

The Boltzmann-Gibbs measure can also be obtained from a maximum entropy argument~\cite{wilson2011entropy, lasota2013chaos}.  The advantage of this view is that the terms in the potential function can be interpreted as economic constraints.  We consider a potential function with three components to develop a baseline model, although more comprehensive presentations are  possible\footnote{We present a simple aggregated model to illustrate the ideas.  A refined model with disaggregation does not change the underlying arguments.}:
 \begin{equation}
 \label{pot_sum}
 \epsilon^{-1} V(x) =  V_\text{Utility}(x) + \kappa V_\text{Cost}(x) + \delta V_\text{Additional}(x),
 \end{equation}
where $\kappa$ is as before and $\delta > 0$ is an additional parameter.  The utility potential describes consumer welfare arising from utility-based choices; the cost potential enforces capacity limits in the system; and the additional potential is a confining term that represents government initiatives, continued investment or a background level of demand.   If we consider a random variable $X \in \mathbb{R}^M$ that is subject to the following equality constraints
\begin{equation}
\begin{aligned}
\label{constr}
\mathbb{E} \bigg[  V_{\text{Utility}}(X) \bigg] &= C_{\text{Utility}}, \\
\mathbb{E} \bigg[  V_{\text{Cost}}(X) \bigg] &= C_{\text{Cost}}, \\
\mathbb{E} \bigg[  V_{\text{Additional}}(X) \bigg] &= C_{\text{Additional}},
\end{aligned}
\end{equation}
with each $C_i \in \mathbb{R}$, then the maximum entropy distribution of $X$ can be written as the Boltzmann-Gibbs measure whose density is given by~\eqref{BoltzmannGibbs} with reference to the potential function in~\eqref{pot_sum}.  We now consider the meaning of each of these constraints in turn.

\subsubsection{Utility potential}
A natural candidate for a utility potential is a measure of consumer welfare.  For example, welfare may be taken to be the area under the demand curve~\cite{williams1977formation}, given by~\eqref{d_eq}, that is equal to the path integral
\begin{equation}
\label{path}
\begin{aligned}
V_{\text{Utility}}(x) &:=  \int_{x_0}^x \big( D_1(x'), \dots, D_M(x') \big) \cdot dx', \\
& = \alpha^{-1} \sum_{i=1}^N O_i \ln \sum_{j=1}^M \exp(U_{ij}(x_j)) + \text{const},
\end{aligned}
\end{equation}
where we have defined the deterministic utility function
\begin{equation}
\label{utility_func}
U_{ij}(x_j) = \alpha x_j - \beta c_{ij}.
\end{equation}
In Appendix~\ref{Appendix:Utility} we show that~\eqref{utility_func} but with $\alpha$ dependent on $i$ is also obtained by seeking a utility function consistent with a singly-constrained model and the path integral in~\eqref{path}.  The log-sum function is commonly used as a welfare measure in the economics literature~\cite{williams1977formation, small1981applied, de2007logsum}.  To make the connection with random utility maximisation explicit, we define the stochastic utility function for a choice being made from origin zone $i$ as
\begin{equation}
\label{s_utility}
\tilde{U}_{ij}(x_j) = U_{ij}(x_j) + \xi_{ij},
\end{equation}
where the $\xi_{ij}$ are independent and identically distributed Gumbel random variables.  Then under the utility maximisation framework, the expected utility attained from a unit flow leaving origin zone $i$ is
\begin{equation}
\mathbb{E}\bigg[\max_{1 \leq j \leq M}\tilde{U}_{ij}(x_j) \bigg] =  \ln \sum_{j=1}^M \exp(U_{ij}(x_j)) + c,
\end{equation}
where $c$ is the Euler-Mascheroni constant~\cite{williams1977formation}.  The utility potential may then be expressed as the expected utility attained from all flows in units of $\alpha$
\begin{equation}
V_{\text{Utility}}(x) = \alpha^{-1} \sum_{i=1}^N O_i \mathbb{E}\bigg[\max_{1 \leq j \leq M}\tilde{U}_{ij}(x_j) \bigg] + \text{const}.
\end{equation}

Two remarks are in order.  First, the scaling factor of $\alpha^{-1}$ is necessary to ensure that the utility potential is non-constant in the limit $\alpha \to 0$.  Second, the tight bounds~\cite{nielsen2016guaranteed}
\begin{equation}
\begin{aligned}
 -\alpha^{-1}\sum_{i=1}^N {O_i} \bigg\{  \max_{1 \leq j \leq M} U_{ij}(x)  + \ln M \bigg\} \leq V_{\text{Utility}}(x)  \leq - \alpha^{-1}\sum_{i=1}^N {O_i} \bigg\{   \max_{1 \leq j \leq M} U_{ij}(x)  \bigg\},
\end{aligned}
\end{equation}
show that $V_{\text{Utility}}(x)$ may be finite when any $x_j \to -\infty$, and so an additional potential is needed to prevent zones from collapsing from a lack of activity.  The bounds again show that the utility potential is closely related to the best alternative available to each of the origin zones.

\subsubsection{Cost potential}
The cost potential prevents each zone from becoming too large, and is justified by the notion that running costs increase with size.  We therefore require that $\lim_{x_j \to + \infty} V(x) = +\infty$.  In view of the equality constraints in~\eqref{constr}, an appropriate cost potential is the total size or capacity of the system
\begin{equation}
\label{cost_pot}
V_{\text{Cost}}(x) = \sum_{j=1}^M W_j(x_j ),
\end{equation}
in which $W_j(x_j) = \exp(x_j)$ is as before.  Since we have
\begin{equation}
\label{cand_pot}
\frac{\partial V_\text{Cost}(x)}{\partial x_j} = W_j(x_j ),
\end{equation}
this choice of potential yields the linear cost term in the overdamped Langevin dynamics considered in~\eqref{lang}.

\subsubsection{Additional potential}
The final potential term must satisfy $\lim_{x_j \to -\infty} V(x) = +\infty$ and must grow sufficiently fast at infinity in order for~\eqref{BoltzmannGibbs} to be well-defined.  The purpose of the additional potential is to prevent zones from collapsing from a lack of activity.  Such mechanisms are commonplace in urban and regional systems, for example, continued investment or government initiatives.  In view of the equality constraints in~\eqref{constr}, a suitable potential function is
\begin{equation}
\label{add_pot}
V_{\text{Additional}}(x) = \sum_{j=1}^M x_j,
\end{equation}
which ensures that the attractiveness of each zone is finite.  The partial derivatives of the additional potential function are
\begin{equation}
\label{cand_pot}
\frac{\partial V_\text{Additional}(x)}{\partial x_j}  = 1.
\end{equation}
Therefore the finiteness constraint requires that there is an additional positive constant term in the deterministic part of the SDE model, given by~\eqref{lang}, to ensure that the SDE has a well-defined stationary distribution.

\subsection{Model Summary}

In summary, we have specified the following potential function 
\begin{equation}
\begin{aligned}
\label{chosen_pot}
\epsilon^{-1} V(x) & =  \underbrace{ -\alpha^{-1} \sum_{i=1}^N O_i  \ln \sum_{j=1}^M \exp(\alpha x_j - \beta c_{ij}) }_{\text{Utility}}  &+ \underbrace{ \kappa \sum_{j=1}^M \exp(x_j )}_{\text{Cost}} - \underbrace{\delta \sum_{j=1}^M x_j}_{\text{Additional}},
\end{aligned}
\end{equation}
which satisfies the assumptions in Appendix~\ref{Appendix:Assumptions}.  The potential function is similar the one obtained by reformulating the Harris and Wilson model in~\eqref{lang}-\eqref{hw_pot}, however, it contains an additional term to prevent zones from collapsing.  A stochastic generalisation of the Harris and Wilson model is given by the overdamped Langevin diffusion in~\eqref{lang}, for which the process converges at a fast rate to the well-defined Boltzmann-Gibbs measure described by~\eqref{BoltzmannGibbs}.  The corresponding size dynamics, in the form of~\eqref{shw}, are given by the Stratonovich SDE
\begin{equation}
\label{stochastic_harris_wilson2}
{dW_j} = \epsilon W_j \bigg (D_j - \kappa W_j + \delta \bigg) dt + \sigma W_j \circ {dB_j}, \qquad W(0) = w_0,
\end{equation}
which is a stochastic generalisation of the original Harris and Wilson model that includes a positive shift to the multiplicative scale factor.  In the limit $\delta, \sigma \to 0$, we obtain the original Harris and Wilson model in~\eqref{HW}.

In the regime $\delta \to 0$, the potential function has stationary points coinciding with the fixed points of the original Harris and Wilson model.  The stationary points for the potential function are given by $M$ simultaneous equations
\begin{equation}
\label{equil}
\sum_{i=1}^N O_i \frac{W_j^\alpha \exp (-\beta c_{ij})}{\sum_{k=1}^M W_k^\alpha\exp(- \beta c_{ik})} = \kappa W_j - \delta, \qquad j=1,\dots,M.
\end{equation}
 Whilst the behaviour of the stochastic and deterministic models may be similar in low-noise regimes over finite time intervals, we emphasise that the asymptotic behaviour differs greatly between the two.  Here we consider a deterministic model to be given by~\eqref{stochastic_harris_wilson2} in the limit $\sigma \to 0$.  For a deterministic model, the dynamics will converge to a stable fixed point satisfying~\eqref{equil}, as determined by the initial condition.  For a stochastic model, the system will converge to a statistical equilibrium that does not depend on the initial condition.  As $t \to +\infty$, the stochastic model spends more time around the lower values of $V(x)$, which occur around stable stationary points, as summarised by the limiting stationary distribution given by~\eqref{BoltzmannGibbs}.
 
\begin{figure}
	\centering
	\includegraphics[scale=.44]{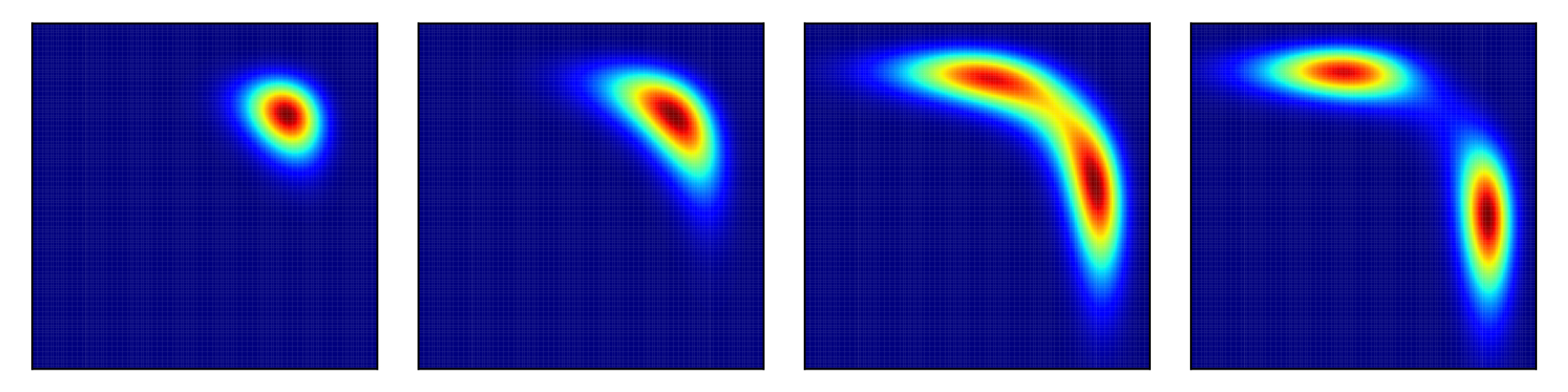}
	\put(-348,95){$\alpha = 0.5$}
	\put(-256,95){$\alpha = 1.0$}
	\put(-158,95){$\alpha = 1.5$}
	\put(-65,95){$\alpha = 2.0$}
		\put(-340,-1){\tiny $X_1$}
	\put(-245,-1){\tiny $X_1$}
	\put(-145,-1){\tiny $X_1$}
	\put(-55,-1){\tiny $X_1$}
	\put(-387,45){\tiny $X_2$}
	\caption{Illustration of the potential function for a small model comprising of two competing zones.  The profile of the potential function is largely determined by the $\alpha$ and $\beta$ pairing; in the illustration above we have held $\beta$ fixed and show $e^{-\gamma V(x)}$ for different values of $\alpha$.}
	\label{pot2d}
\end{figure}
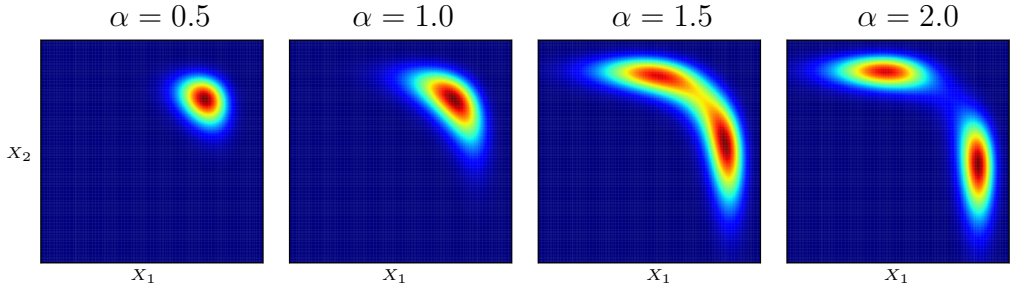

 We now comment on the Boltzmann-Gibbs measure described by~\eqref{BoltzmannGibbs}.  The Boltzmann-Gibbs measure is the equilibrium distribution of~\eqref{lang}, but is also justified as a probability distribution for urban and regional structure with a maximum entropy argument.  When considering the Boltzmann-Gibbs measure, we specify $\epsilon=1$ to avoid over-parametrising the model since the relative level of noise is controlled by the inverse temperature $\gamma = 2\sigma^{-2}$.  As $\gamma \to +\infty$, the Boltzmann distribution collapses to a Dirac mass around the global minimum of $V(x)$, which is unlikely to provide a good fit to the observed urban structure.  As $\gamma \to 0$, the distribution of sizes approaches an improper uniform distribution.  The profile of $V(x)$ is largely influenced by the pair of $\alpha$ and $\beta$ values, as illustrated in Figure~\ref{pot2d}.  A large $\alpha$ relative to $\beta$ results in all activity taking place in one of the zones, whereas this regime is unlikely when $\alpha$ is low relative to $\beta$.

 Lastly, we can use the deterministic model to specify appropriate values of the cost of floorspace $\kappa$ and the additional parameter $\delta$.  By defining
\begin{equation}
\label{kappa}
\kappa = \frac{1}{K} \Bigg(  \sum_{i=1}^N O_i + \delta M \Bigg),
\end{equation}
the deterministic model converges to an equilibrium with a total size of $K$ units.  Setting $\kappa$ as in~\eqref{kappa} is justified with a supply and demand argument~\cite{dearden2011framework}.  For simplicity, we use $K=1$; the choice is arbitrary.  We can then specify $\delta$ relative to the size of the smallest zone possible, since at equilibrium the size of a zone with no inward flows is $\delta / \kappa$.

\section{Parameter estimation}
\label{S:hier}
In this section we consider the inverse problem; the task of determining $\alpha$ and $\beta$ from observed urban structure.  The value of $\alpha$ describes consumer preference towards more popular destinations and the value of $\beta$ describes how much consumers are inconvenienced by travel.  We use retail activity as an archetype, however, our methodology is general and can be applied to other singly-constrained systems. Whilst $\alpha$ and $\beta$ can be estimated using discrete choice models~\cite{mcfadden1978modeling, mcfadden1981econometric, mcfadden2000mixed, anas1983discrete}, this approach requires large volumes of flow data and is impractical for large systems.  We instead make use of the model described by the Boltzmann-Gibbs measure in~\eqref{S:p_meas_urban_systems}.

We formulate the task of inversion as a statistical inference problem, as advocated in~\cite{kaipio2006statistical}.  The Bayesian approach is based on the following principles: the unknown parameters are modelled as random variables; our degree of uncertainty is described by probability distributions; and the solution to the inverse problem is the posterior probability distribution.  Unlike classical methods, a Bayesian approach is well-posed and allows us to incorporate prior knowledge of the unknowns into the modelling process. A Bayesian approach yields a posterior probability over the model parameters, and the parameter values can be determined from the posterior mean or maximum a posteriori estimates.

\subsection{A Bayesian approach to parameter estimation}
The Boltzmann distribution in~\eqref{S:p_meas_urban_systems} is assumed to form an integral part of the data generating process, however, further uncertainty arises from measurement noise\footnote{It may be necessary to include a model error term if the Boltzmann-Gibbs measure provides a poor fit to the data.}.  To this end, we assume that an observed a configuration of urban structure $Y \in \mathbb{R}_{>0}^M$ is related to some latent sizes $W \in \mathbb{R}_{>0}^M$ and multiplicative noise $E \in \mathbb{R}_{>0}^M$ by
\begin{equation}
\label{random_like}
\ln Y = \ln W + \ln E.
\end{equation}
Multiplicative noise is appropriate since all measurements are positive and there is more scope for error when measuring larger zones.  As before, it is natural to work in terms of log-sizes $X = \ln W \in \mathbb{R}^M$ and we assume that $X \sim \rho_\infty$ is a realisation of the Boltzmann-Gibbs measure given by~\eqref{BoltzmannGibbs} and~\eqref{chosen_pot}.  The latent variables $X$ depend on the model parameters $\Theta = \{ \alpha, \beta \} \in \mathbb{R}^2_{>0}$, which we summarise by a single variable for notational convenience.  We assume that $\ln E \sim N(0, \Sigma)$ is a realisation of Gaussian noise for some symmetric positive definite covariance matrix $\Sigma \in \mathbb{R}^{M \times M}$.  

We specify a prior $\pi(\theta)$ on the model parameters.  The prior distribution for the latent variables is denoted $\pi(x | \theta)$ and is given by~\eqref{BoltzmannGibbs}, which we repeat here to make the $\theta$-dependence explicit in our notation
\begin{equation}
\pi(x | \theta) = \frac{1}{z(\theta)}\exp\big( -\gamma V_\theta(x) \big), \qquad z(\theta) = \int_{\mathbb{R}^M} \exp\big( -\gamma V_\theta(x) \big) dx.
\end{equation}
We emphasise that $\pi(x | \theta)$ is only known up to a normalizing constant $z(\theta)$ that is a function of $\theta$.  The likelihood function $\pi(y | x)$ is the Gaussian density given by~\eqref{random_like}.  The joint posterior density then has the form
\begin{equation}
\label{joint_post}
\pi(x, {\theta} | y) \propto \pi(\theta) \frac{1}{z(\theta)} \exp\big( -\gamma V_\theta(x) \big) \pi(y | x),
\end{equation}
and is `doubly-intractable' as both the normalisation factor of~\eqref{joint_post} and the function $z(\theta)$ are unknown.  The estimation of $z(\theta)$ is a notoriously challenging problem as it requires the integration of a complex function over a high-dimensional space ~\cite{ gelman2014bayesian, murray2006}.  The normalization constant $z(\theta)$ is a probability-weighted sum of all possible outcomes and is a necessary penalty against model complexity. The $\theta$-dependence for the $z(\theta)$-term poses significant computational challenges as the joint posterior density cannot be evaluated at all, not even up to an irrelevant multiplicative constant.

\subsection{Computational strategies}
To explore the posterior distribution we resort to numerical simulation and use Markov chain Monte Carlo (MCMC) to estimate integrals of the form
\begin{equation}
\label{obj}
\mathbb{E}\big[g(X, \Theta) | Y = y \big] := \int_{\mathbb{R}^2_{>0}} \int_{\mathbb{R}^M} g(x, \theta) \pi(x, \theta | y) dx d\theta,
\end{equation}
where $g(x, \theta)$ is an integrable function of interest.  For example,~\eqref{obj} can be used to compute the mean, variance and density estimates of the posterior marginals.  As suggested in~\cite{gelman2014bayesian}, we can use an approximate method to estimate $z(\theta)$.   We consider the quadratic approximation of $V_\theta(x)$ that is obtained from a second-order Taylor expansion around its global minima $m_\theta$
\begin{equation}
\label{quad_approx}
\hat{V}_\theta(x) =	V_\theta(m_\theta) + \frac{1}{2}(x - m_\theta)^T \Delta V_\theta(m_\theta) (x - m_\theta).
\end{equation}
Since the integral for $z(\theta)$ only has significant contributions in the neighbourhood of $m_\theta$, where~\eqref{quad_approx} is a good approximation, we estimate $z(\theta)$ as
\begin{equation}
\begin{aligned}
\label{zapprox}
z(\theta) &\approx \int_{\mathbb{R}^M} \exp \big( {-\gamma \hat{V}_\theta(x)} \big) dx, \\ &= \exp \big( {-\gamma V_\theta(m_\theta)} \big) \int_{\mathbb{R}^M} \exp \bigg( {-\frac{\gamma}{2}(x - m_\theta)^T \Delta V_\theta(m_\theta) (x - m_\theta)} \bigg) dx, \\
& = \exp \big( {-\gamma V_\theta(m_\theta)} \big) \frac{{ \big(2 \pi \gamma^{-1}\big)^{M/2}} }{\big|  \Delta V_\theta(m_\theta) \big|^{1/2}}.
\end{aligned}
\end{equation}
This is known as a saddle point approximation and is asymptomatically accurate as $\gamma \to +\infty$~\cite{butler2007saddlepoint}.  In all but special cases, the global minima of $V_\theta(x)$ is unique and can be found inexpensively using Newton-based optimisation, for example, using the L-BFGS algorithm with the right initial condition~\cite{nocedal2006numerical}.  We run the optimisation procedure for multiple initialisations to provide a good coverage of the basins, although this is only necessary for $\alpha > 1$.  The curvature term $\Delta V_\theta(m_\theta)$ is given by~\eqref{hessian}.  With~\eqref{zapprox} we can proceed with the MCMC scheme in Appendix~\ref{Appendix:MCMC}.

To obtain more accurate posterior summaries, especially in the case that the saddle point approximation performs poorly, we look towards a consistent estimator of~\eqref{obj}.  Despite the intractable $z(\theta)$ term, we are able to construct a Markov Chain $\big\{ X^{(i)}, \Theta^{(i)}, \Omega^{(i)} \big\}_{i=1}^n$ such that
\begin{equation}
\label{obj_est}
\mathbb{E}\big[g(X, \Theta) | Y = y \big] =  \lim_{n \to + \infty} \frac{\sum_{i=1}^n \Omega^{(i)} g\big(X^{(i)}, \Theta^{(i)}\big)}{\sum_{k=1}^n \Omega^{(k)} }.
\end{equation}
The estimator requires that we can obtain unbiased estimates of the reciprocal normalizing constant $1/z(\theta)$,  which can obtained by randomly truncating an infinite series involving importance sampling estimates of $z(\theta)$~\cite{murray2006, lyne2015russian}.  The estimator given by~\eqref{obj_est} is an importance-sampling style estimator, but with each weight $\Omega^{(i)} \in \{-1, 1\}$ equal to the sign of the unbiased estimate of $1/z(\theta)$ for that iteration.  The suitability of the scheme is dependent on being able to obtain precise importance sampling estimates of $z(\theta)$, which is challenging for low-noise regimes due to the concentration of measure.  Negative values of $\Omega^{(i)}$ arise from imprecision in the $z(\theta)$ estimates and have the effect of increasing the variance of the estimator given by~\eqref{obj_est}.  Further details of the scheme are in Appendix~\ref{Appendix:MCMC}.

\subsection{Implementation details}
\label{sec:imp}
We specify weakly-informative uniform priors on $\alpha$ and $\beta$, restricted to the interval $[0, 2]$ with a suitable scaling of $\beta$ determined by a preliminary study, as done in~\cite{dearden2015explorations, dearden2011framework}\footnote{In our implementation, we normalize the cost matrix so that all elements sum to $7 \times 10^5$.}.  In this setting, we are able to compare our inferred $\alpha$ and $\beta$ values with the $R$-squared analysis performed for the deterministic Harris and Wilson model in~\cite{dearden2015explorations, dearden2011framework}.  Whilst ideally we would place priors on all parameters that specify the Boltzmann-Gibbs measure, we acknowledge that in doing so we would encounter both identifiability issues and tuning difficulties with regards to the importance sampling scheme for $z(\theta)$.  We are able to proceed by fixing the remaining hyperparameters to suitable values.  We specify $\epsilon = 1$ to avoid over-parametrising the model and specify $\gamma$ to reflect a desired level of noise.  We set $\delta$ to the size of the smallest zone.  This is justified by considering the Gamma distribution of a zone with no inward flows.  We normalize the origin quantities and total sizes to determine $\kappa$ from~\eqref{kappa}.   Lastly,  For demonstration purposes, we specify independent and homogeneous observation noise by setting $\Sigma = \lambda^2 I$ where $\lambda$ is the standard deviation of the noise.

To compute low-order summary statistics of the form in~\eqref{obj}, we use the Monte Carlo scheme in Appendix~\ref{Appendix:MCMC}, comprising of a block Gibbs scheme.  We use a Metropolis-Hastings random walk with reflective boundaries for the $\Theta$-updates and Hamiltonian Monte Carlo (HMC) for the $X$-updates.  We tune the step size parameter for the $\Theta$-updates to obtain an acceptance rate in the range $30\%-70\%$, and we tune the step size and number of steps for the $X$-updates to obtain an acceptance rate of at least $90\%$.   For the $\Theta$-updates, we either require global minima of $V_\theta(x)$, for the saddle point approximation in~\eqref{zapprox}, or unbiased estimates of $1/z(\theta)$, for the pseudo-marginal MCMC framework described in Appendix~\ref{Appendix:MCMC}.  When requiring a global minima of $V_\theta(x)$, we perform multiple runs of the L-BFGS algorithm for $M$ different initial conditions. When requiring consistent estimates of $1/z(\theta)$, we use annealed importance sampling (AIS) with HMC transition kernels.  We initialize AIS with the log-gamma distribution that can be obtained from $\pi(x | \theta)$ by letting $\alpha, \beta \to 0$.    We produce unbiased estimates by truncating an infinite series of importance sampling estimates with a random stopping time $T$ with $\text{Pr}(T \geq k) \propto k^{-1.1}$, and therefore requiring $T+1$ runs of AIS.  Running AIS a large number of times is a computationally intensive task, however, the estimates can be obtained in parallel.

\section{Case study: the London Retail system}

\label{S:case}
In this section, we illustrate our proposed methodology with an aggregate retail model using London data similar to the example in~\cite{dearden2015explorations, dearden2011framework}\footnote{The code used for the results in this section is available at
	\url{https://github.com/lellam/cities_and_regions}}.  Whilst the model can be improved with disaggregation to capture further problem-specific characteristics, the underlying arguments would remain the same.  We demonstrate how the Boltzmann-Gibbs measure can be used to simulate configurations of urban structure before setting out to infer the $\alpha$ and $\beta$ values in the utility function.  In the context of retail, the attractiveness term in~\eqref{utility_func} is justified by the benefit consumers gain from the improved range of options and economies of scale, and the cost term represents inconvenience of travel.  The inverse problem is of particular interest in the context of retail as the flow data is difficult to obtain.  On the other hand, urban structure is relatively straight-forward to measure and may be routinely available.    Whilst some attempts have been made in the literature to estimate the parameters of a similar spatial interaction model~\cite{dearden2015explorations, dearden2011framework}, these approaches are somewhat ad hoc but do provide a basis of comparison.

\begin{figure}
	\includegraphics[scale=1.]{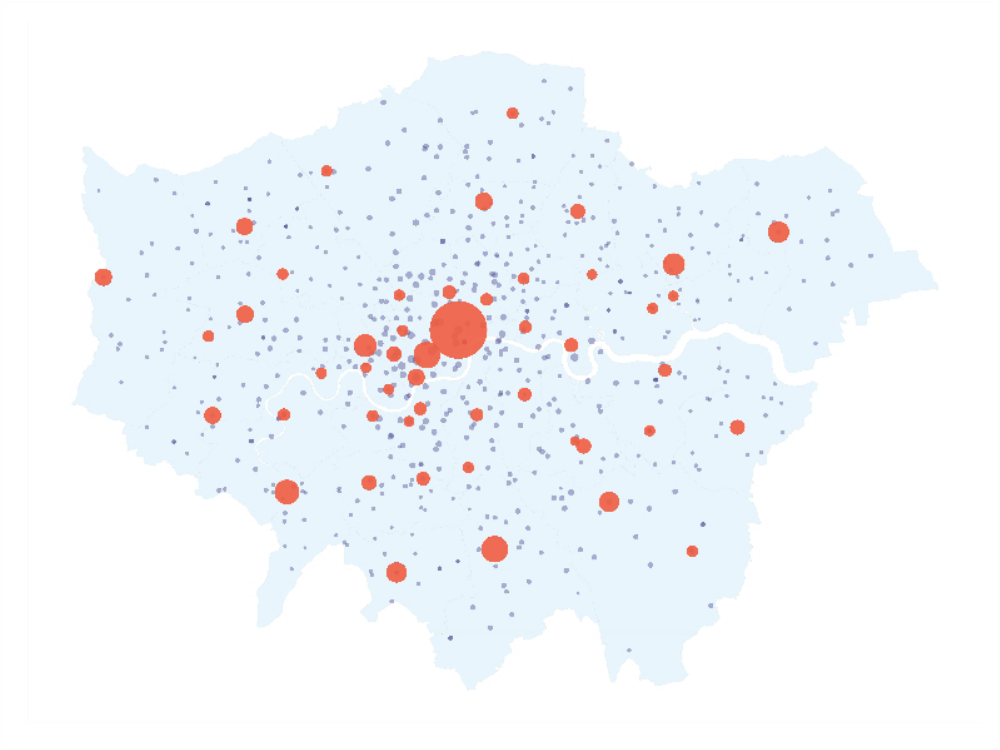}
	\centering
	\caption{Visualisation of the observation data $Y$ (red) over the map of London.  The red markers indicate the destination zones, which are the $49$ town centres, and the blue markers indicate the origin zones, which are the $625$ residential wards.  The sizes of the markers are given by the respective $Y$ and $O$ values, and each zone is plotted at its longitude-latitude coordinate.}
	\label{data}
\end{figure}

We obtain measurements of retail floorspace for London town centres for $2008$ from a London Town Centre Health Check report prepared by the Greater London Authority~\cite{london20092009}.  We only include town centres classified as international, metropolitan and major town centre classifications in our study, giving $M=49$ town centres.  The remaining town centres are mostly district town centres that have a relatively high concentration of convenience goods and more localised catchment; we argue that these would be better modelled separately.  We determine the origin quantities from ward-level household and income estimates, with $N=625$, published by the Greater London Authority~\cite{authority2016london1, authority2016london2}.  We take the origin quantities to be the spending powers as given by the population size multiplied by the average income.    The floorspace measurements and residential data is presented in Figure~\ref{data}, over the map of London~\cite{authority2016london3}.  In our implementation, we calculate the cost matrix from Euclidean distance, although a better representation would use a transport network~\cite{dearden2015explorations}.

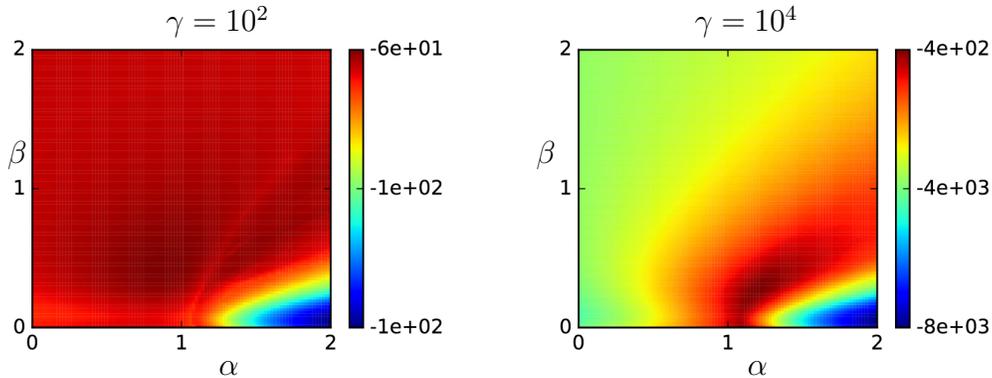
\begin{figure}
	\includegraphics[scale=0.62]{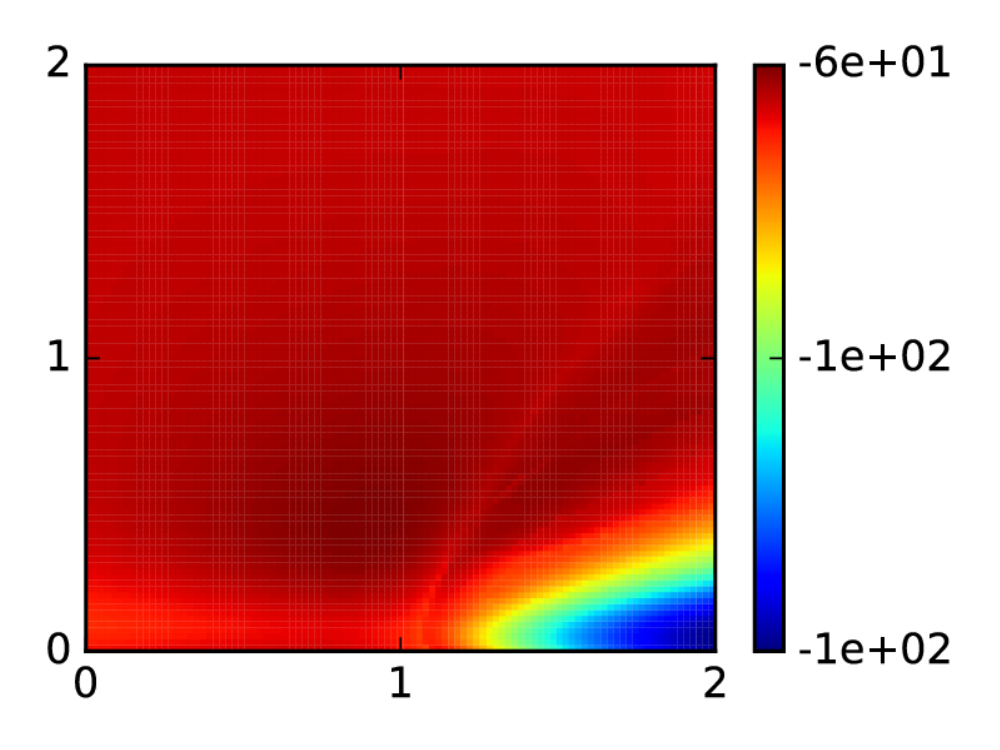} \qquad
	\includegraphics[scale=0.62]{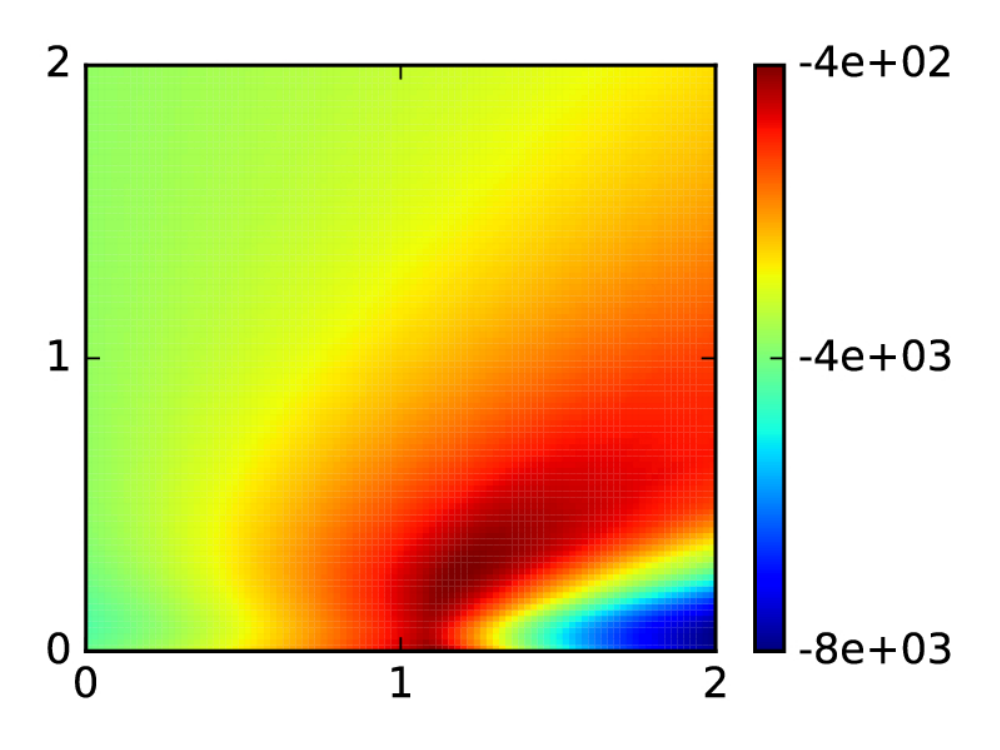}
	\put(-300, 0){$\alpha$}
	\put(-320, 130){$\gamma = 10^2$}
	\put(-120, 130){$\gamma = 10^4$}
	\put(-100, 0){$\alpha$}
	\put(-380, 80){$\beta$}
	\put(-180, 80){$\beta$}
	\centering
	\caption{Evaluations of the logarithm of~\eqref{test} over a grid of $100 \times 100$ values of $\alpha$ and $\beta$ for a regime with high-noise (left) and a regime with low-noise (right).}
	\label{opt_results}
\end{figure}

We first perform a preliminary study of our model in the limit of no observation noise $\lambda \to 0$, in which case the $\theta$-marginal of~\eqref{joint_post} is
\begin{equation}
\label{test}
\begin{aligned}
\pi( {\theta} | y) \propto \pi(\theta) \frac{1 }{z(\theta)} \exp \big( -\gamma V_\theta(x) \big).
\end{aligned}
\end{equation}
With this simplification, we are able to evaluate the posterior probabilities over a grid of $\alpha$ and $\beta$ values.  We evaluate the probabilities over a $100 \times 100$ grid for $\gamma = 10^2$ and $\gamma = 10^4$, representing high-noise and low-noise regimes, respectively.  Using the justification given in the previous section, we specify $\delta=0.006$ and $\kappa = 1.3$.   We produce the grid by estimating $z(\theta)$ with the saddle point approximation in~\eqref{zapprox}\footnote{It is very difficult to estimate $z(\theta)$ for high values of $\gamma$ using importance sampling techniques due to the concentration of measure.}. The results are presented in Figure~\ref{opt_results}, in which the scales indicate that the model with high-noise provides the better explanation of the data.  We find that the best fit for the high-noise regime is $\alpha=0.90$ and $\beta=0.46$ and the best fit for the low-noise regime is $\alpha=1.18$ and $\beta=0.28$.  As expected, the low-noise regime suggests stronger attractiveness effects as the model with a higher level of noise is able to explain variation by stochastic growth.  The alpha and beta values are positively correlated; this can be seen in Figure~\ref{opt_results} and is due to the competing effects in the utility function in~\eqref{utility_func}.

In~\cite{dearden2015explorations}, the authors perform an $R$-squared analysis that we replicate here for our deterministic version of the Harris and Wilson model for a basis of comparison.  The predicted value $W_\text{Pred}$ is taken to be the equilibrium obtained from the ODE model given by~\eqref{stochastic_harris_wilson2} with $\sigma \to 0$ and the initial condition $w_0 = Y$.  The $R$-squared value is defined as $R^2 = 1 - {SS}_\text{res}/{SS}_\text{tot}$, where ${SS}_\text{res}/{SS}_\text{tot}$ is the ratio of the variance of the residuals $Y-W_\text{Pred}$ and the variance of the observed $Y$.  Whilst our Bayesian approach is fundamentally different, and we should not expect to obtain too similar results, the $R$-squared analysis yield a best fit of $\alpha=1.36$ and $\beta=0.42$.  This is agreeable with the findings for the low-noise regime in Figure~\ref{opt_results}.  Furthermore, there are some strong similarities between the profile of posterior probabilities and the profile of the $R$-squared values.  First, both approaches find that the poorest fit is for a regime in which $\alpha$ is too high and beta is too low; these values result in most activity taking place in a single zone.  Second, both approaches agree in that a good fit can be found for $1 < \alpha < 2$.

\begin{figure}
	\centering
	\includegraphics[scale=.29]{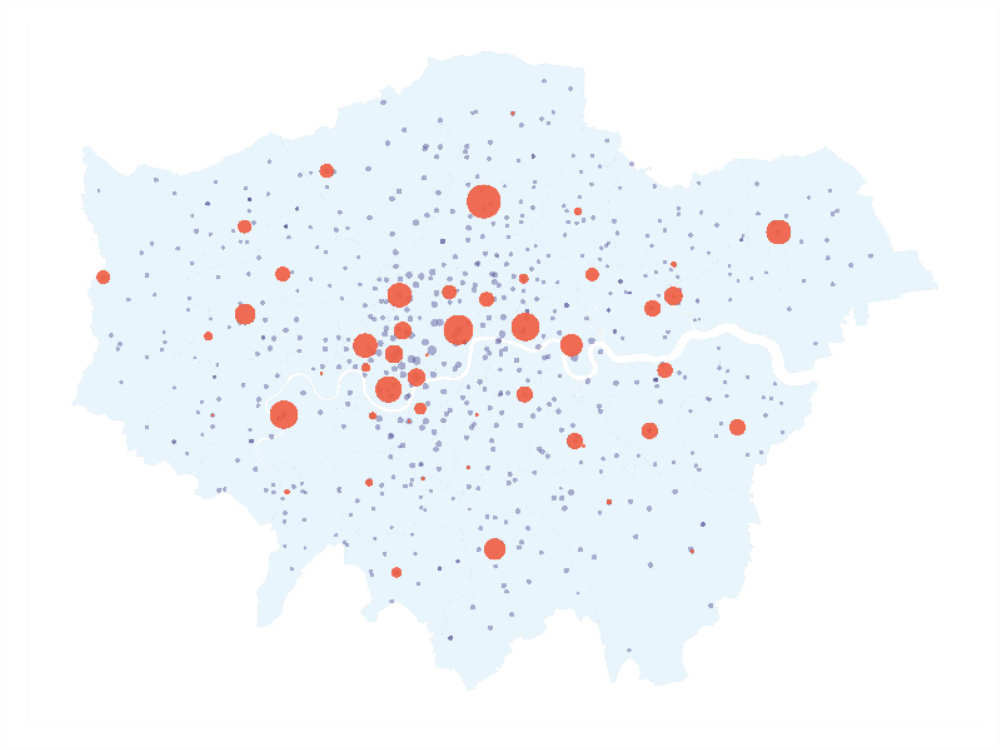}
	\includegraphics[scale=.29]{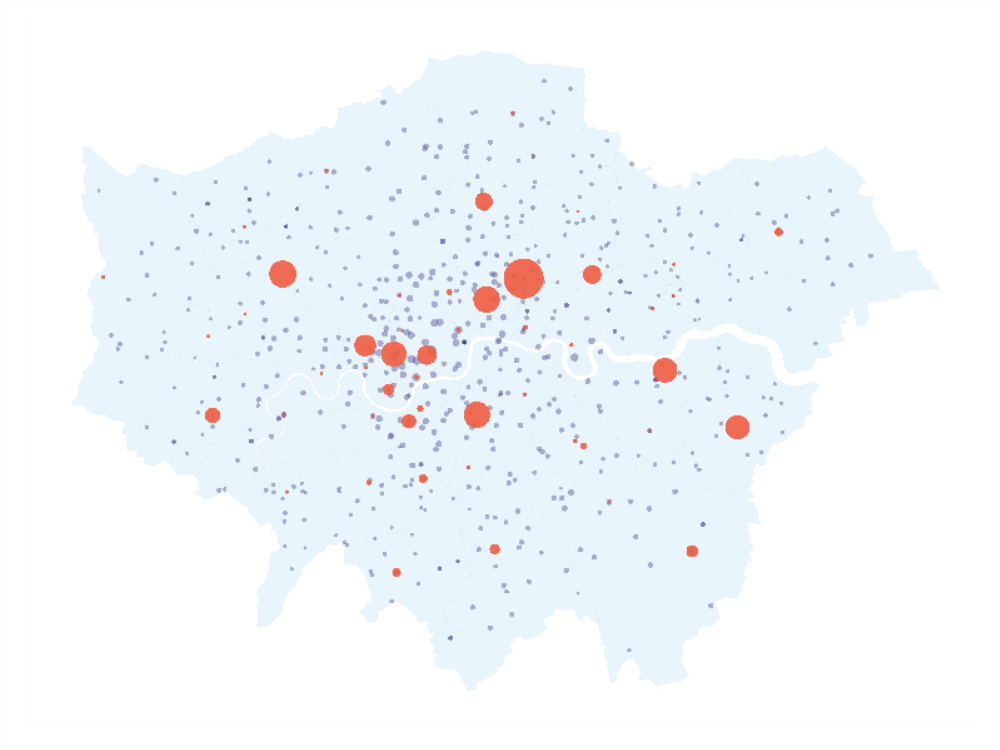} 
	\includegraphics[scale=.29]{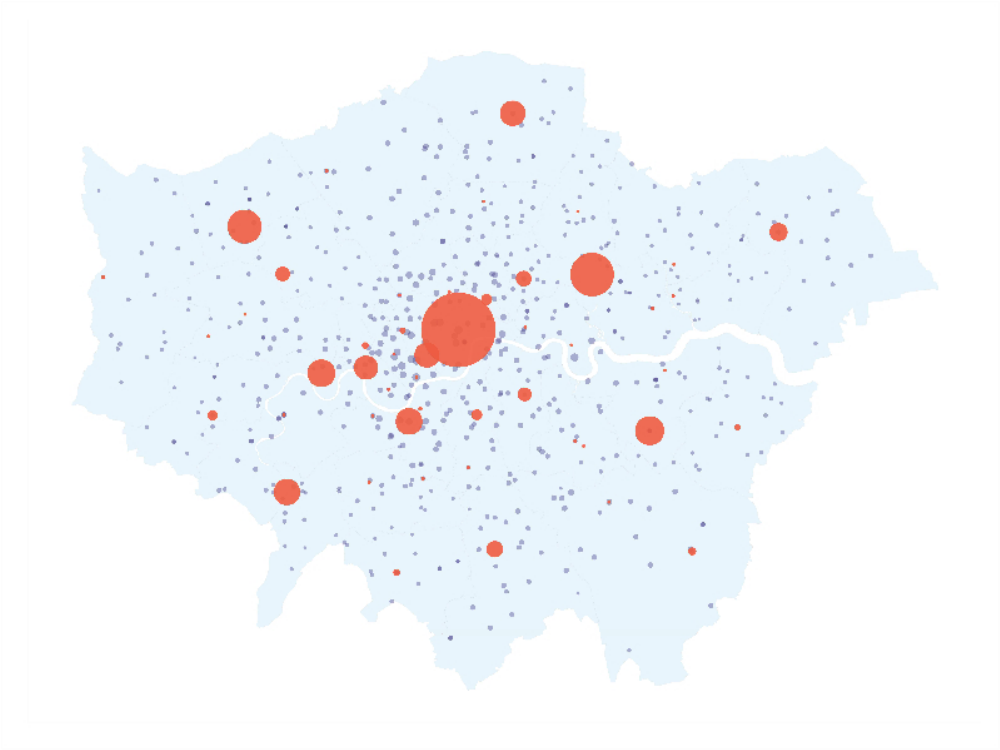} 
	\includegraphics[scale=.29]{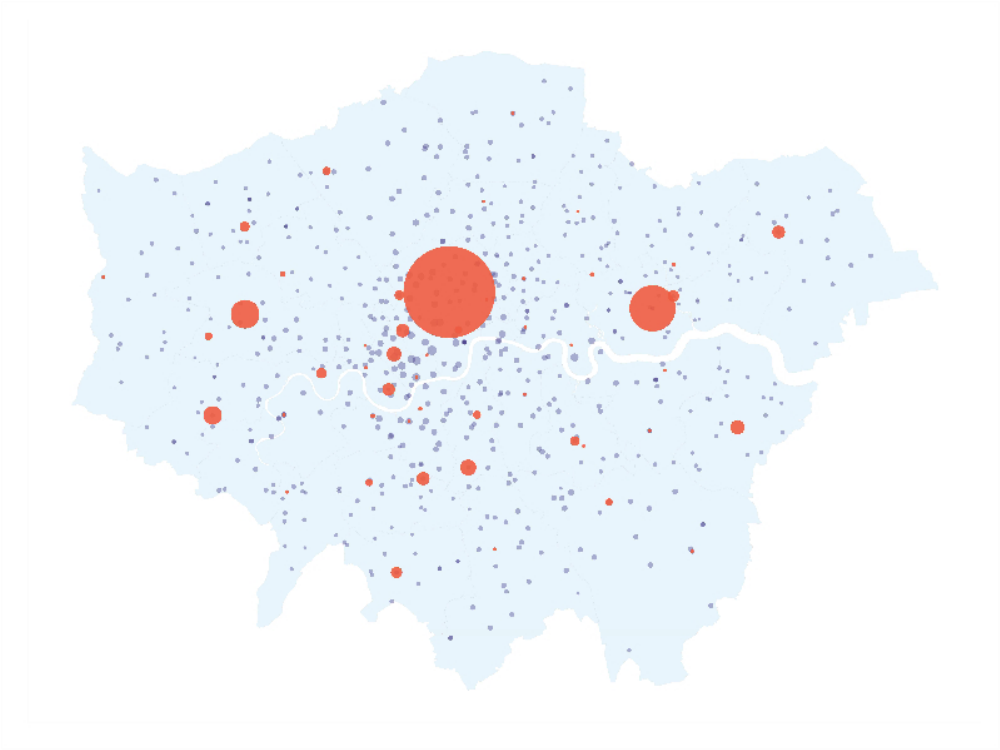} 
	\put(-370,30){Draw 1}
	\put(-315, 70){$\alpha = 0.5$}
	\put(-230,70){$\alpha = 1.0$}
	\put(-145,70){$\alpha = 1.5$}
	\put(-60,70){$\alpha = 2.0$} \\
	\includegraphics[scale=.29]{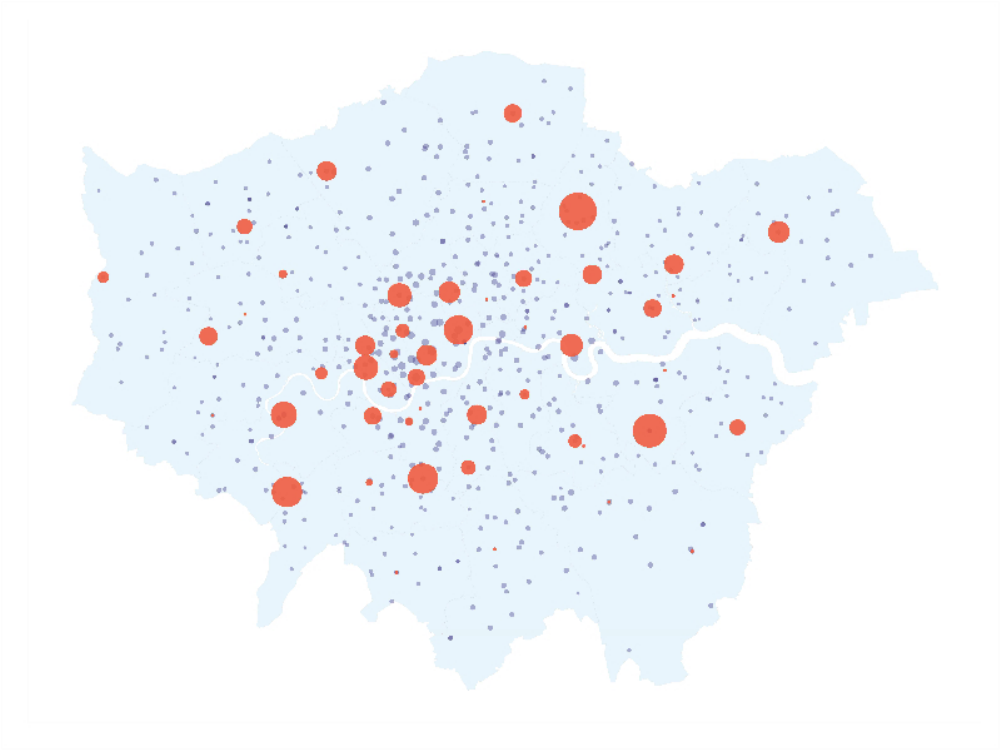}
	\includegraphics[scale=.29]{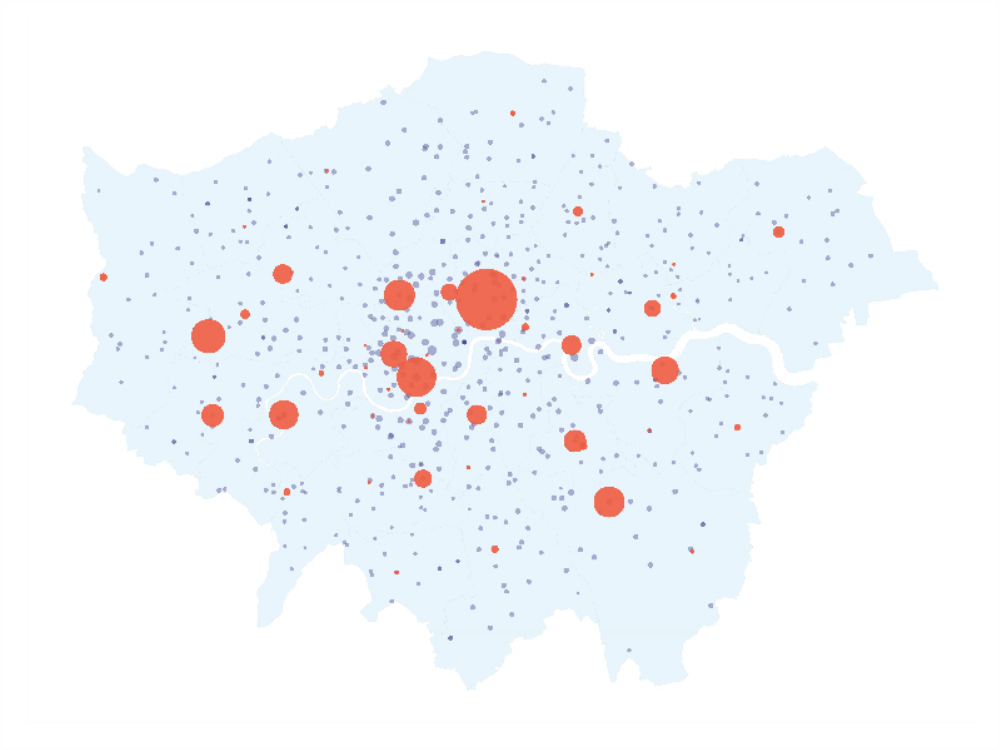} 
	\includegraphics[scale=.29]{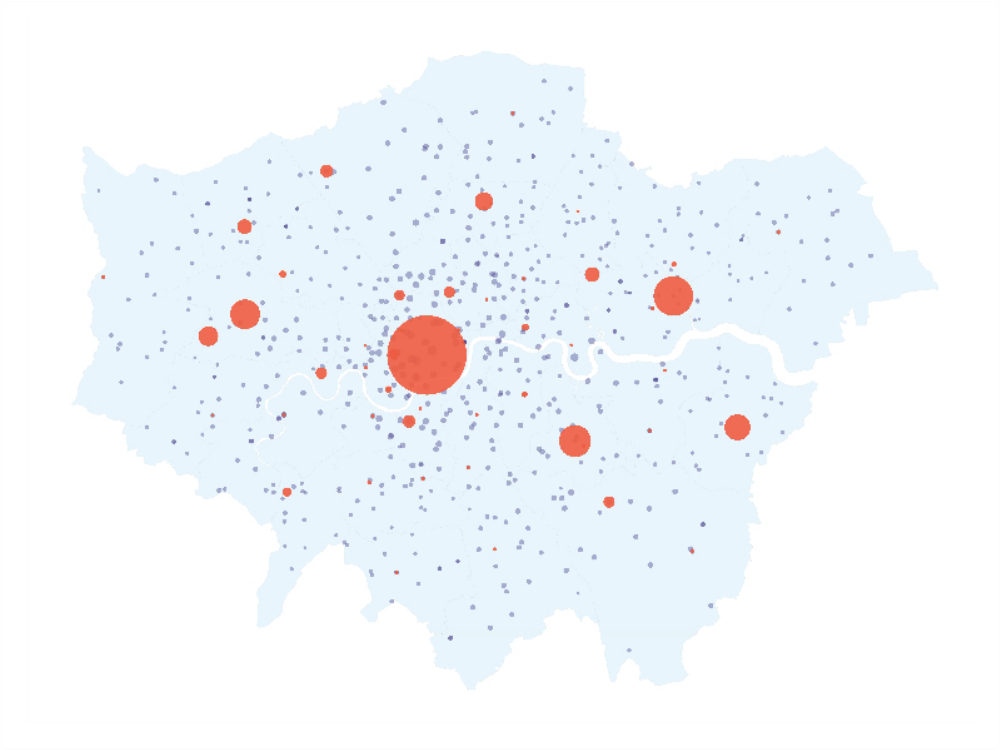} 
	\includegraphics[scale=.29]{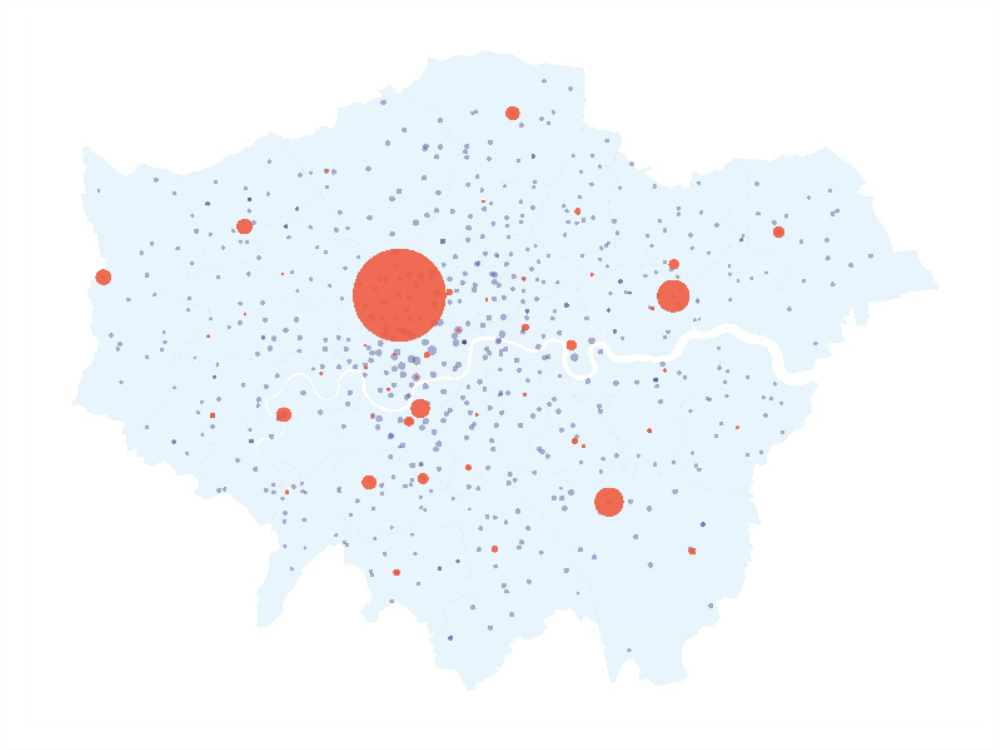} 
	\put(-370,30){Draw 2} \\
	\includegraphics[scale=.29]{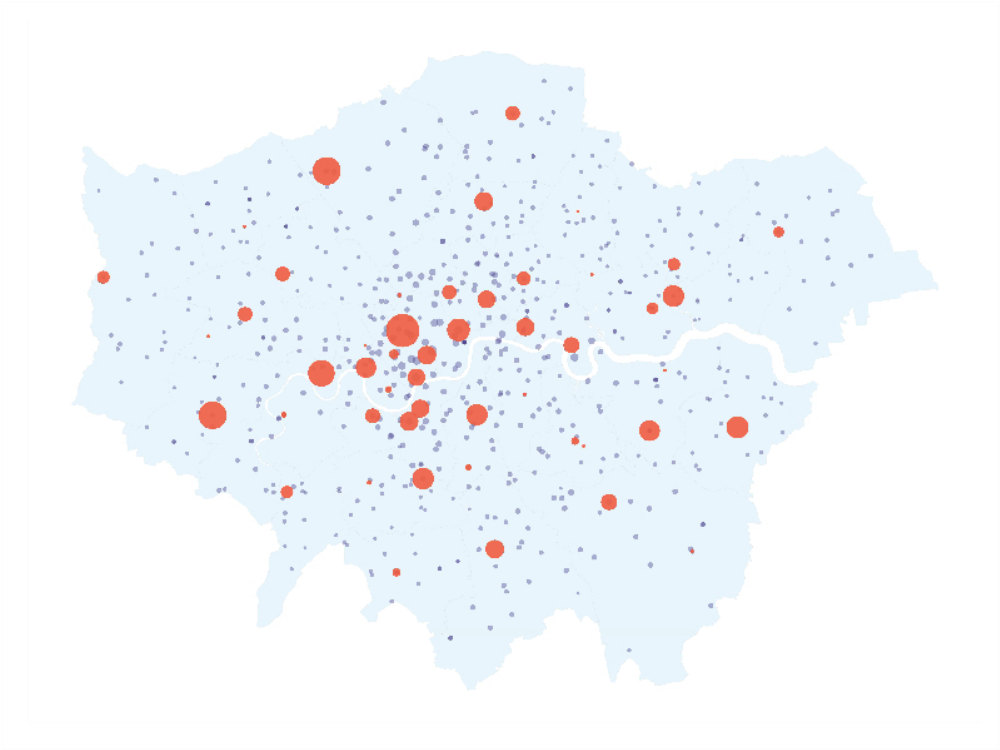}
	\includegraphics[scale=.29]{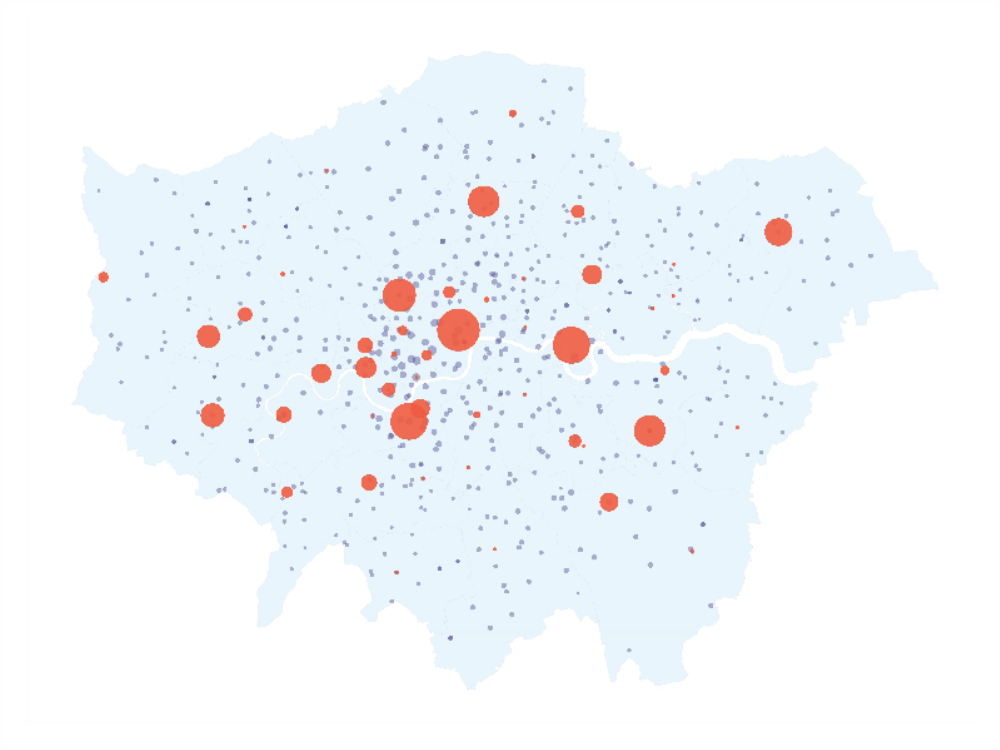}
	\includegraphics[scale=.29]{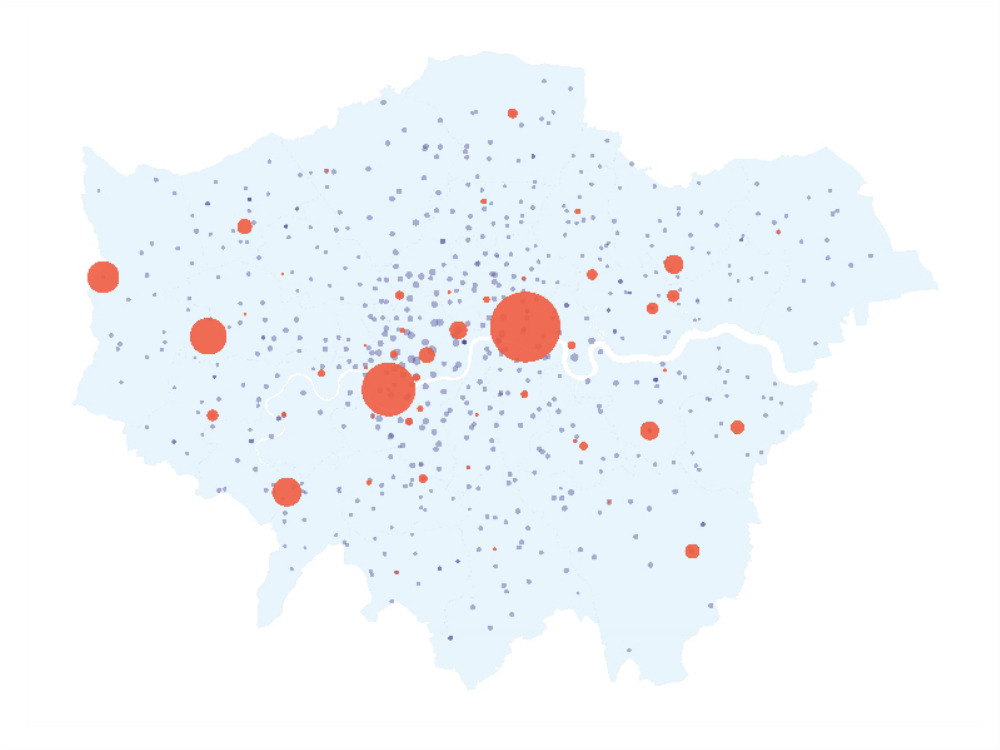}
	\includegraphics[scale=.29]{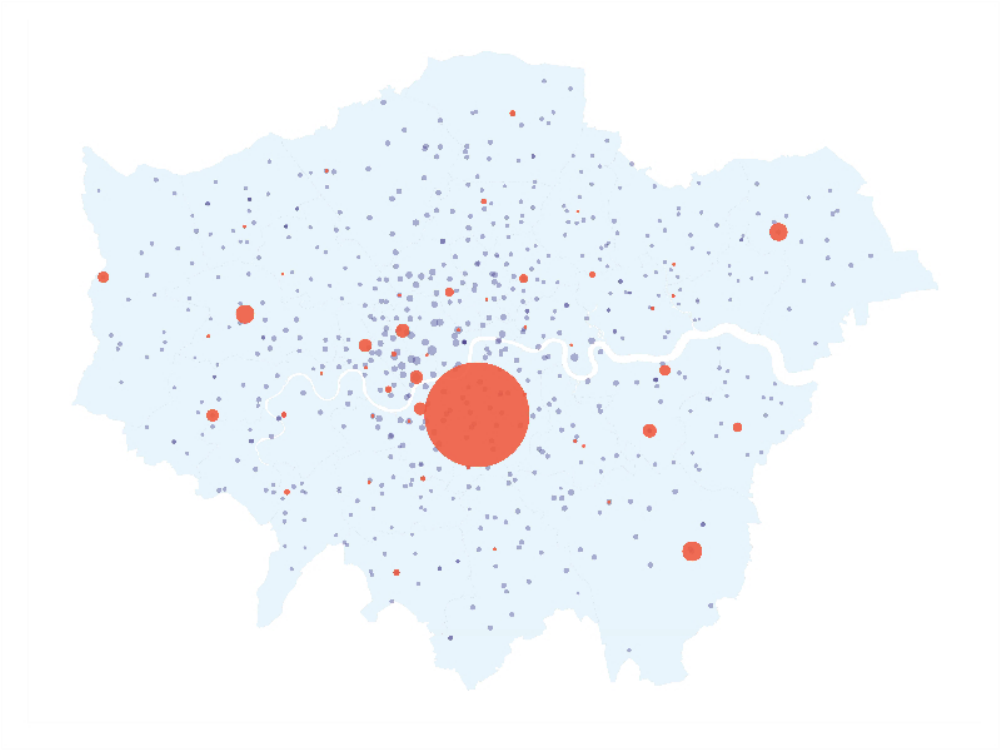} 
	\put(-370,30){Draw 3} \\
	\includegraphics[scale=.29]{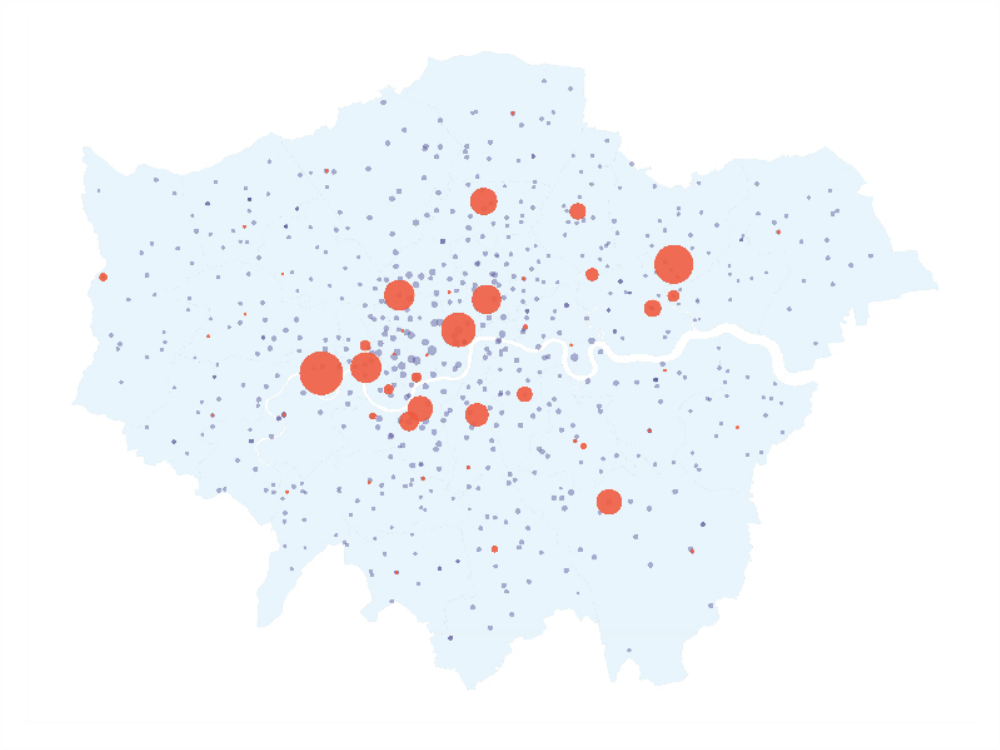}
	\includegraphics[scale=.29]{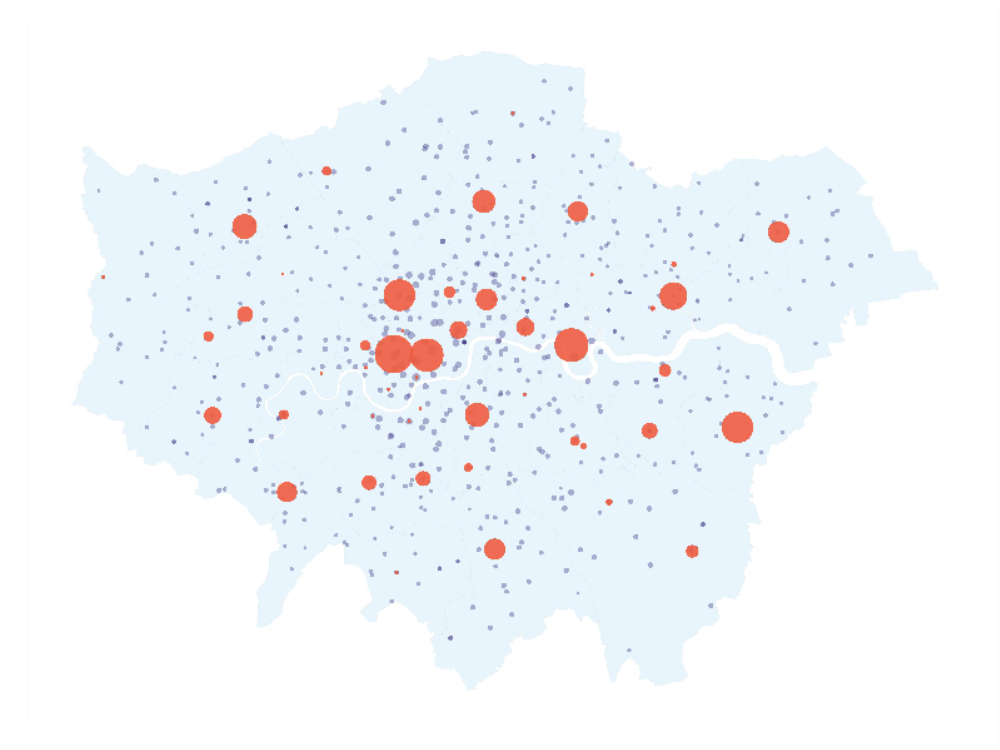}
	\includegraphics[scale=.29]{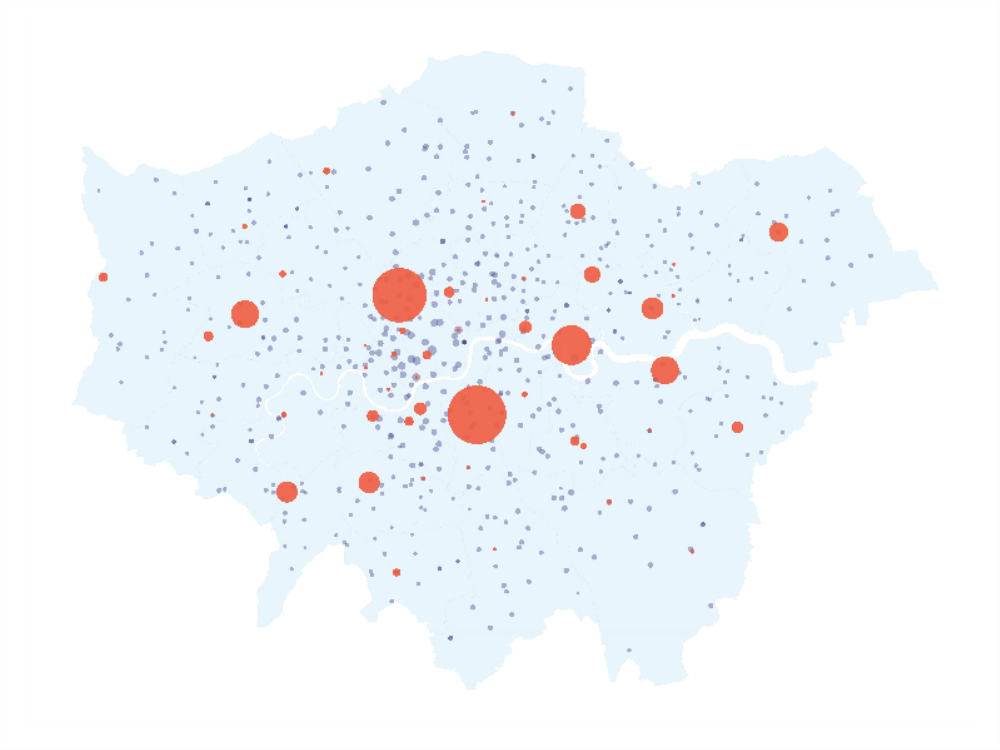}
	\includegraphics[scale=.29]{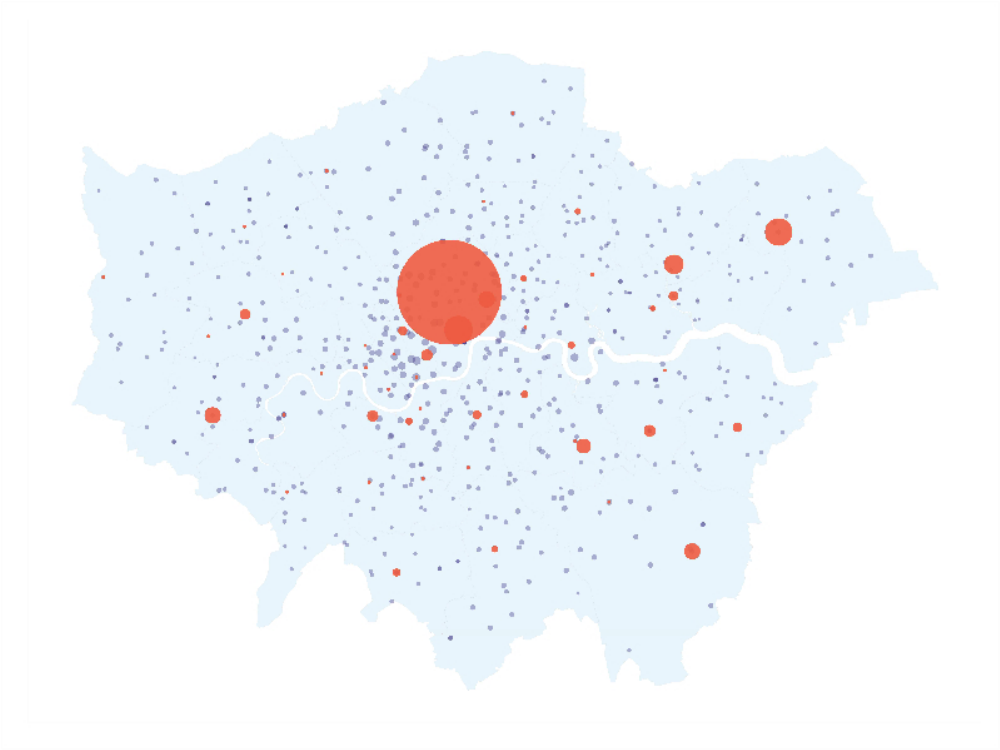} 
	\put(-370,30){Draw 4} \\
	\caption{Approximate draws of the latent variables from  $\pi(x | \theta)$ for a high-noise regime with $\gamma=10^2$, obtained by running a Markov chain of length $10,000$ using HMC combined with parallel tempering.  Each row shows $4$ randomly selected states from the Markov chain with $\alpha$ as specified and $\beta = 0.5$.}
	\label{fig:lon_prior}
\end{figure}

\begin{figure}
	\centering
	\includegraphics[scale=.29]{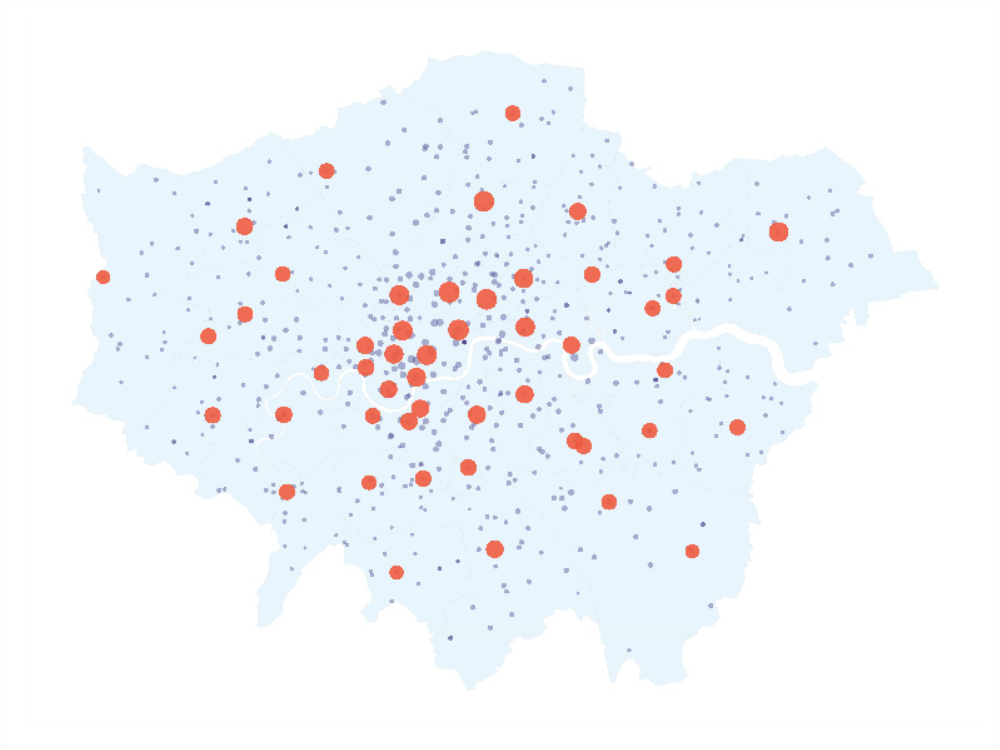} 
	\includegraphics[scale=.29]{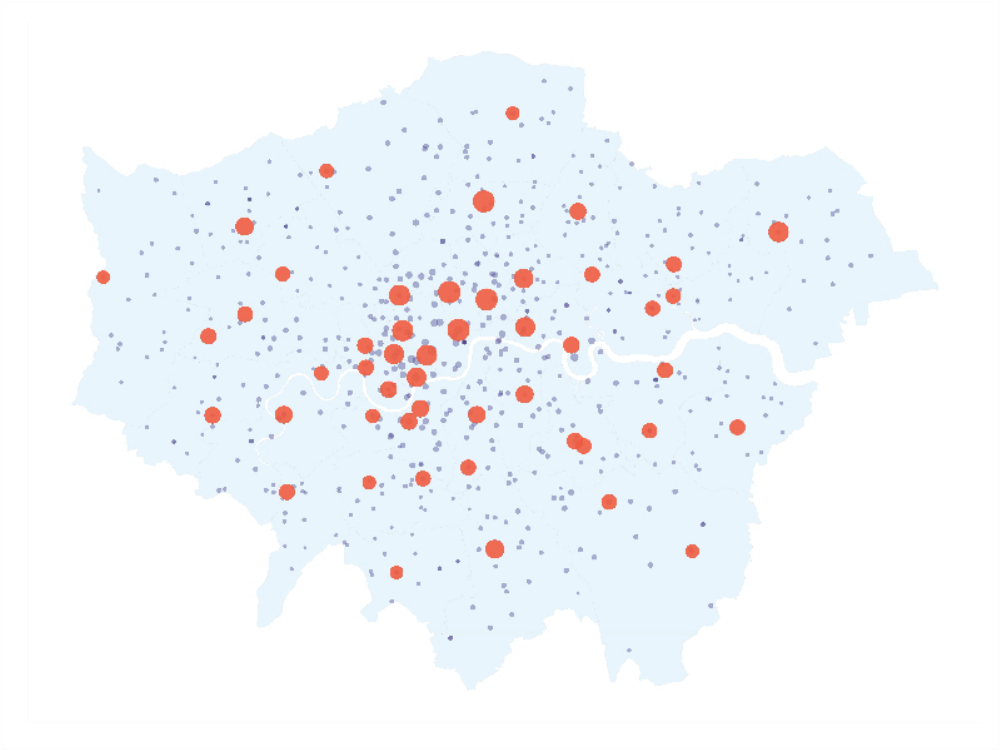} 
	\includegraphics[scale=.29]{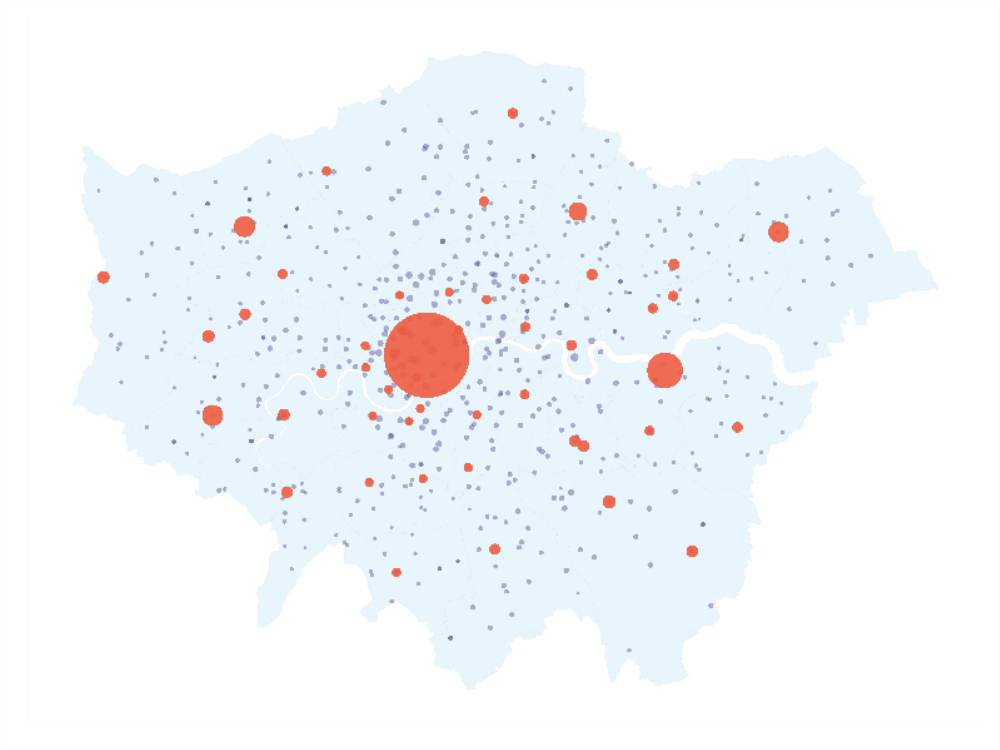} 
	\includegraphics[scale=.29]{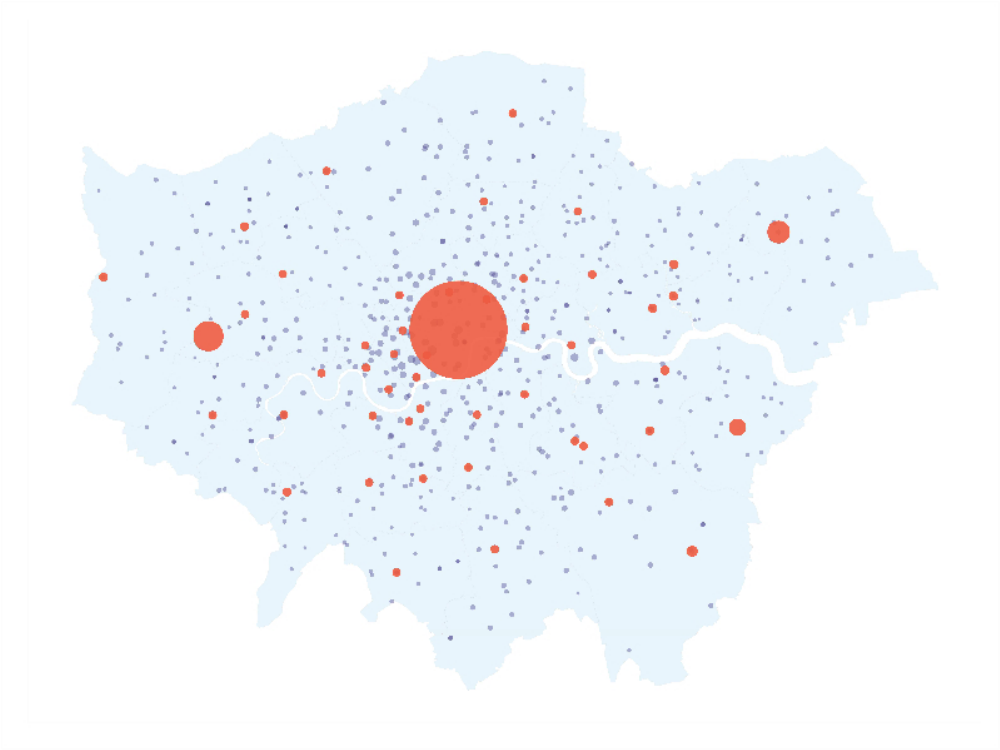} 
	\put(-315, 70){$\alpha = 0.5$}
	\put(-230,70){$\alpha = 1.0$}
	\put(-145,70){$\alpha = 1.5$}
	\put(-60,70){$\alpha = 2.0$} \\
	\caption{Global minima of the latent variables from  $\pi(x | \theta)$, obtained by running the L-BFGS algorithm for $M$ different initial conditions.  These configurations are representative of draws from $\pi(x | \theta)$ in a low-noise regime with $\gamma \gg 1$.}
	\label{fig:lon_prior2}
\end{figure}

Next, we draw the latent variables from the prior distribution $\pi(x | \theta)$ to verify the suitability of the modelling.  For illustrative purposes, we consider a range of $\alpha$ values across $[0, 2]$, and hold $\beta=0.5$ fixed. For the regime with high-noise, the approximate draws are obtained by running a Markov chain of length $10,000$ using HMC combined with parallel tempering for $5$ different temperature levels~\cite{liu2008monte}.  For the regime with low-noise, we plot configurations of the global minima of $V_\theta(x)$ obtained from numerical optimisation as there is little variation between samples. The results are in Figures.~\ref{fig:lon_prior}-\ref{fig:lon_prior2}, respectively.  It can be seen that higher values of $\alpha$ and lower values of $\beta$ create a sparse structure in that all activity takes place in very few zones.  Conversely, lower values of $\alpha$ and higher values of $\beta$ lead to a  more dense structure.

We now return to the observation model in~\eqref{random_like} to account for observation noise in the data.  For illustrative purposes we specify $\lambda = 0.1$ so that the relative noise for a zone of size $1/M$\footnote{Relative noise in this context is $\lambda/ \log M$.} is $3\%$.  Although an improved specification of observation noise from a preliminary study would lead to more accurate inferences, the arguments and methodology we are presenting would remain the same.  We run a Markov chains of length $20,000$.  For the high-noise regime, we use the pseudo-marginal MCMC methodology in Appendix~\ref{Appendix:MCMC}.  Our importance sampling estimates comprised of $10$ particles and $50$ equally spaced inverse-temperatures.  For the low-noise regime, we were unable to obtain precise importance sampling estimates due to the concentration of measure, so used the MCMC methodology in Appendix~\ref{Appendix:MCMC} with the saddle point approximation in~\eqref{zapprox}.  For both examples, the empirical autocorrelation for $\alpha$ and $\beta$ is below $0.2$ after $25$ steps. For pseudo-marginal MCMC scheme, 88\% of the signs are positive, which is acceptably high for the scheme to be used.

\begin{figure}
	\includegraphics[scale=.6]{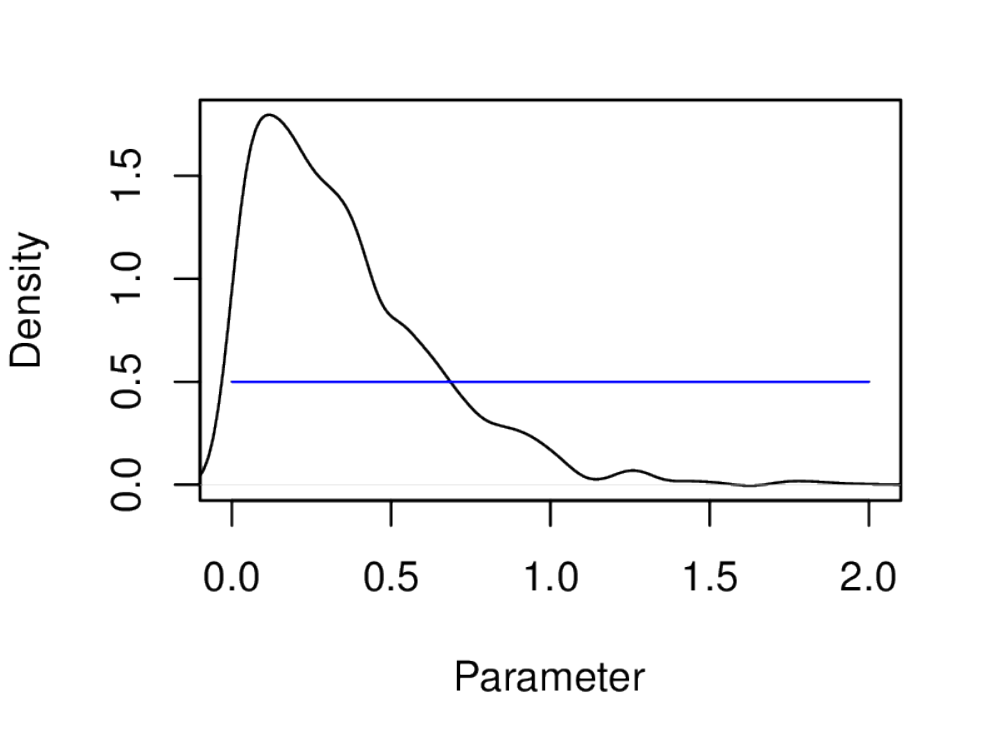} 
	\includegraphics[scale=.6]{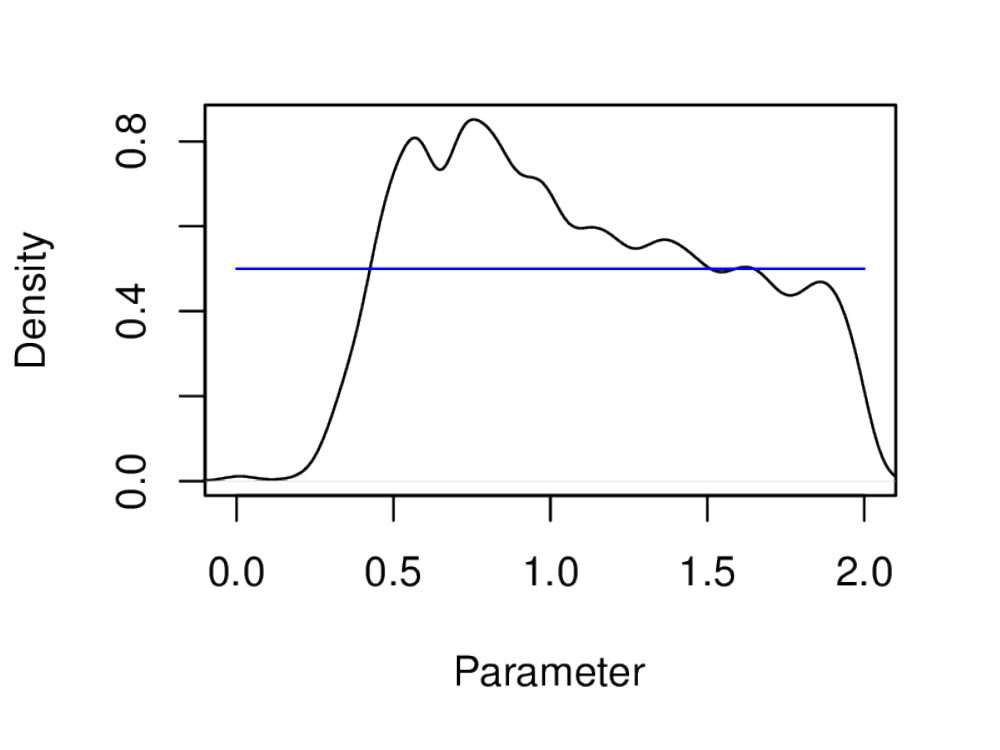} 
	\put(-260,117){\large $\alpha$}
	\put(-85,117){\large $\beta$}
	\centering
	\caption{Posterior marginal density estimates for $\alpha$ and $\beta$ for the high-noise regime ($\gamma = 10^2$).  The smooth density estimates were obtained by applying~\eqref{obj} to a Gaussian kernel. The blue line indicates the uniform prior density.}
	\label{high-density}
\end{figure}

\begin{figure}
	\centering
	\begin{minipage}{0.68\textwidth}
		 \centering
\includegraphics[width=1.\textwidth]{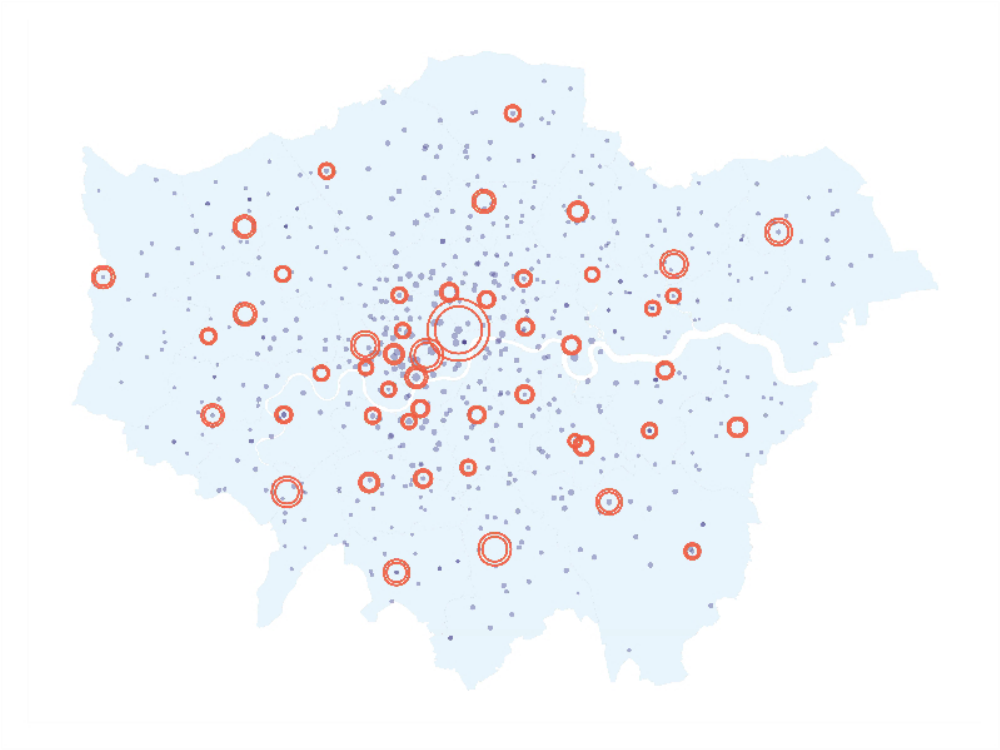}
\end{minipage}\hfill
 \begin{minipage}{0.32\textwidth}
	\includegraphics[width=1.\textwidth]{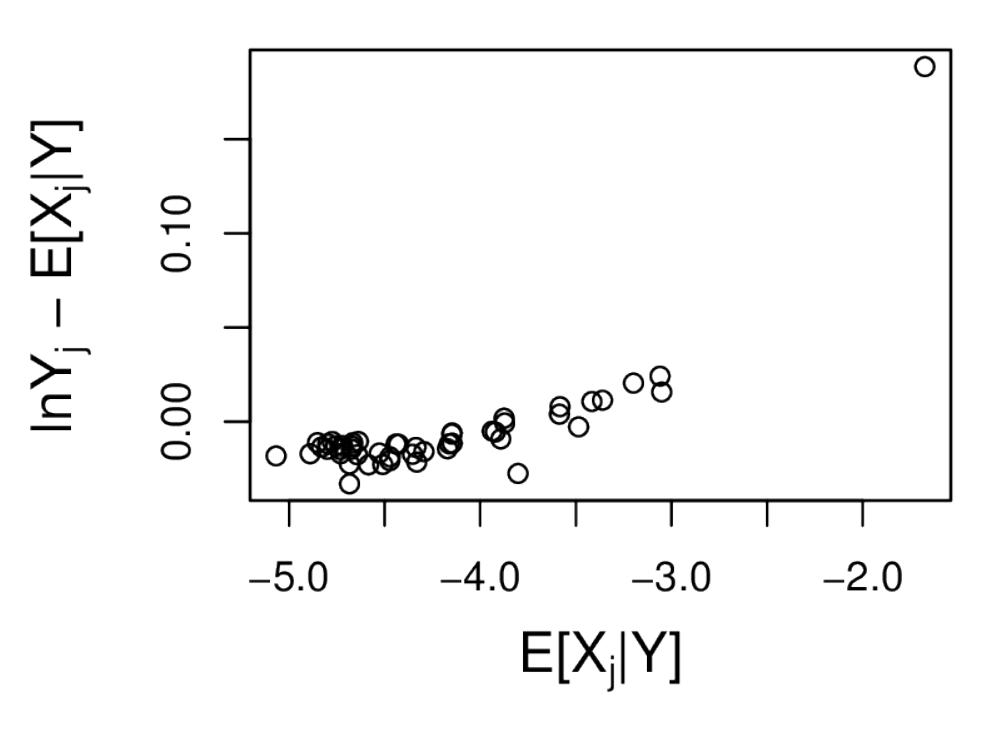}
	\includegraphics[width=1.\textwidth]{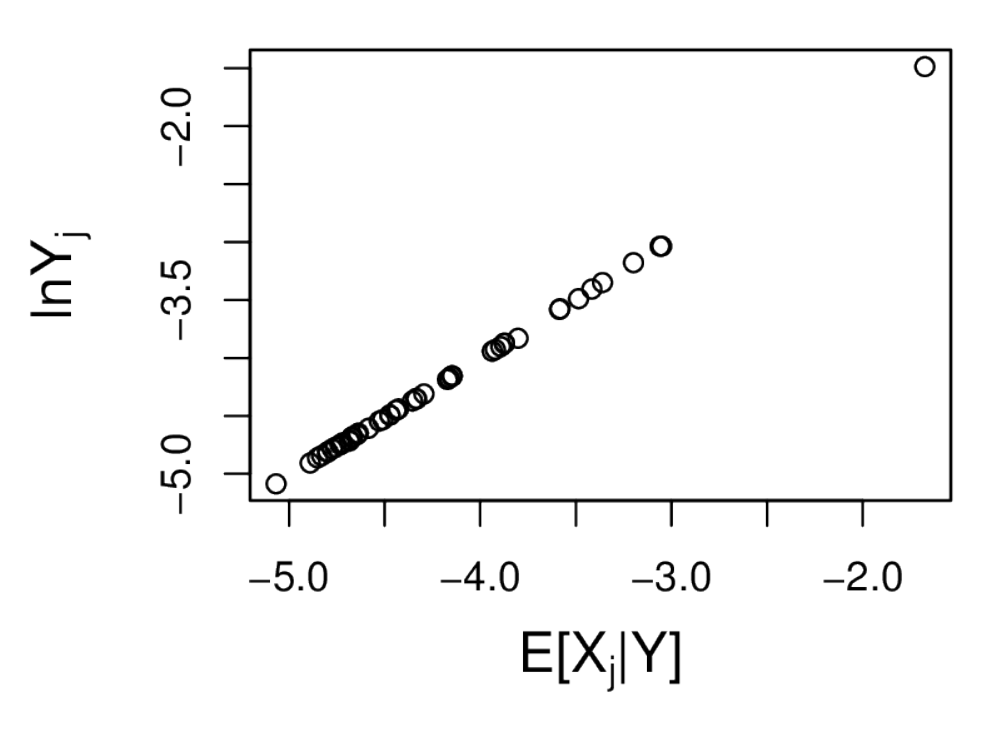}
	   \end{minipage}
\caption{Visualisation of the posterior latent variables $X$ for the high-noise regime ($\gamma=10^2$). Left:  The outer and inner rings show the posterior mean plus or minus three standard deviations, respectively.  Top-right: Expected attractiveness against the expected residual.  Bottom-right: Expected attractiveness against observed value.}
\label{high-lat}
\end{figure}

Plots of the smoothed density estimates for $\alpha$ and $\beta$ for the high-noise regime are presented in Figure~\ref{high-density}.  Plots of the latent sizes showing the posterior mean plus or minus three standard deviations are presented in Figure~\ref{high-lat} alongside plots of the expected residuals and observation data.  The posterior-marginals of $\alpha$ and $\beta$ give mean plus or minus one standard deviation estimates of $0.35 \pm 0.28$ and $1.09 \pm 0.46$, respectively, which appear reasonable in light of the analysis in Figure~\ref{opt_results}.   The plots of the expected residuals and observation data suggest that the model provides a reasonable fit to the data, and that the assumption of homogeneous observation noise is reasonable.  This is to be expected as the high-noise model provides a flexible model. After taking into account the observation noise, a weaker attractiveness effect was observed.

\begin{figure}
\includegraphics[scale=.6]{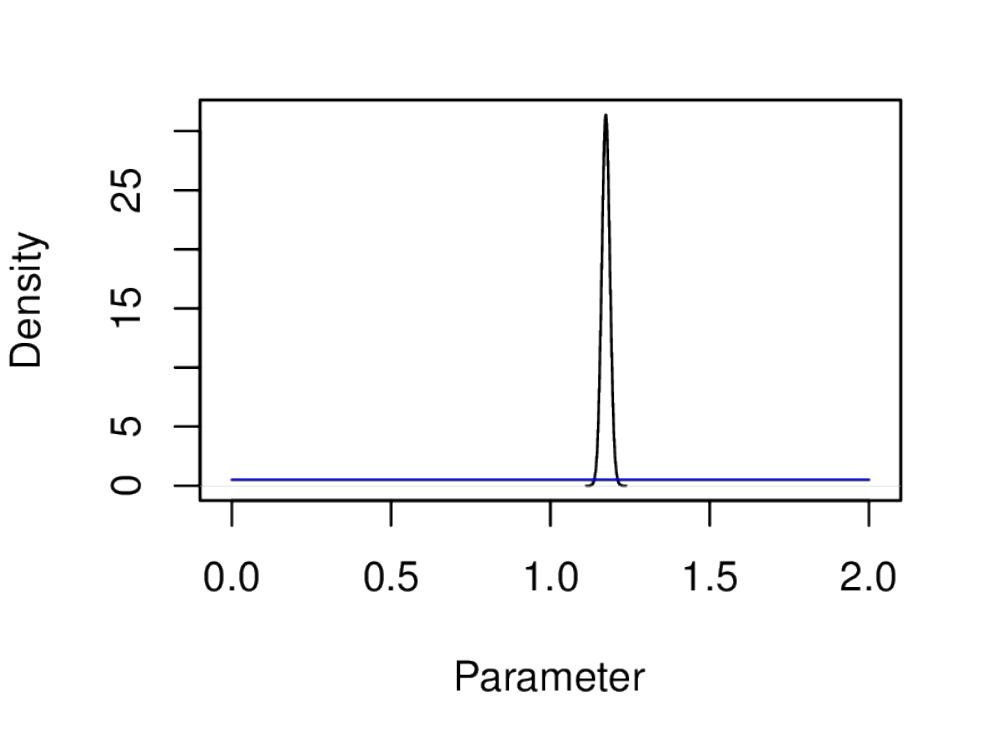} 
\includegraphics[scale=.6]{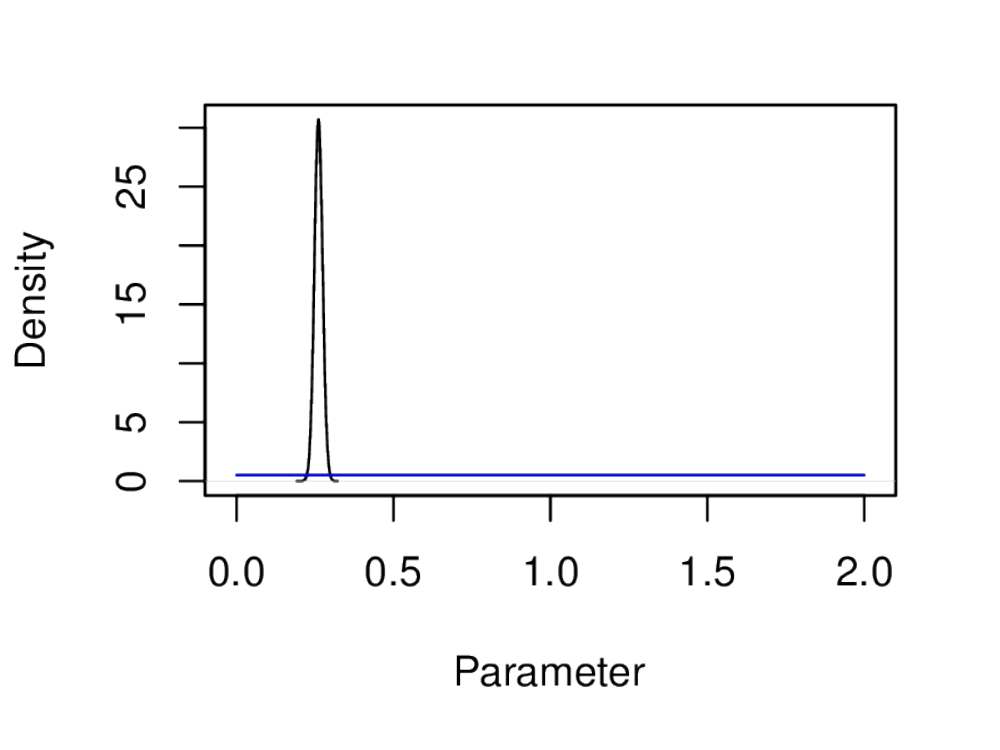} 
\put(-260,117){\large $\alpha$}
\put(-85,117){\large $\beta$}
	\centering
	\caption{Posterior marginal density estimates for $\alpha$ and $\beta$ for the low-noise regime ($\gamma = 10^4$).  The smooth density estimates were obtained by applying~\eqref{obj} to a Gaussian kernel.  The blue line indicates the uniform prior density.}
	\label{low-density}
\end{figure}

\begin{figure}
	\centering
	\begin{minipage}{0.68\textwidth}
		\centering
		\includegraphics[width=1.\textwidth]{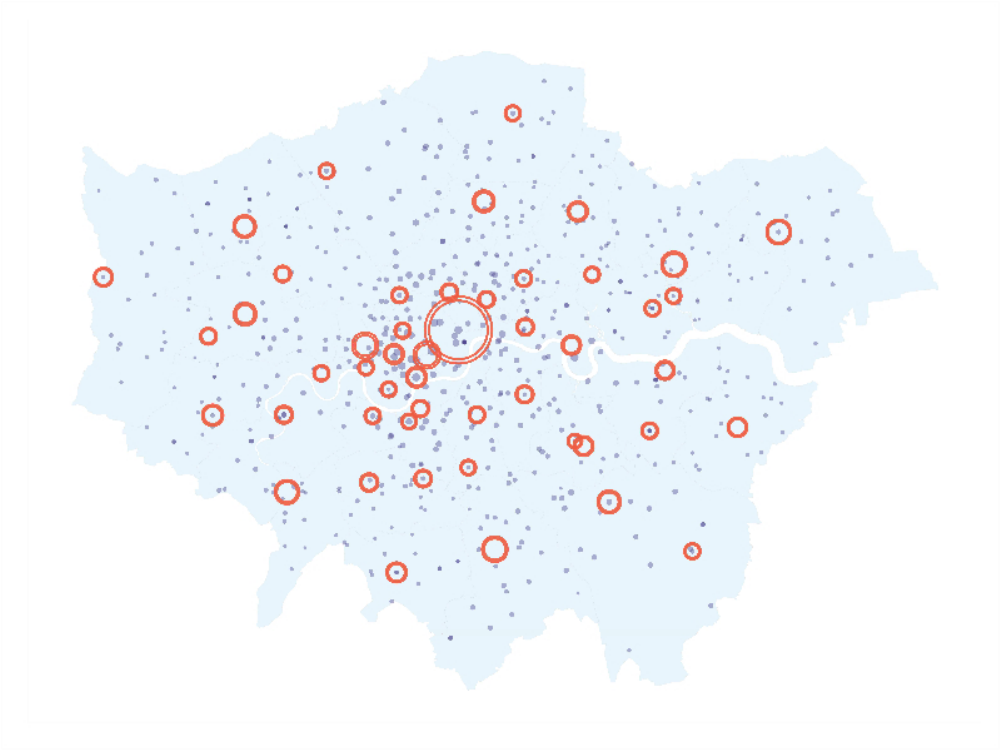}
	\end{minipage}\hfill
	\begin{minipage}{0.32\textwidth}
	\includegraphics[width=1.\textwidth]{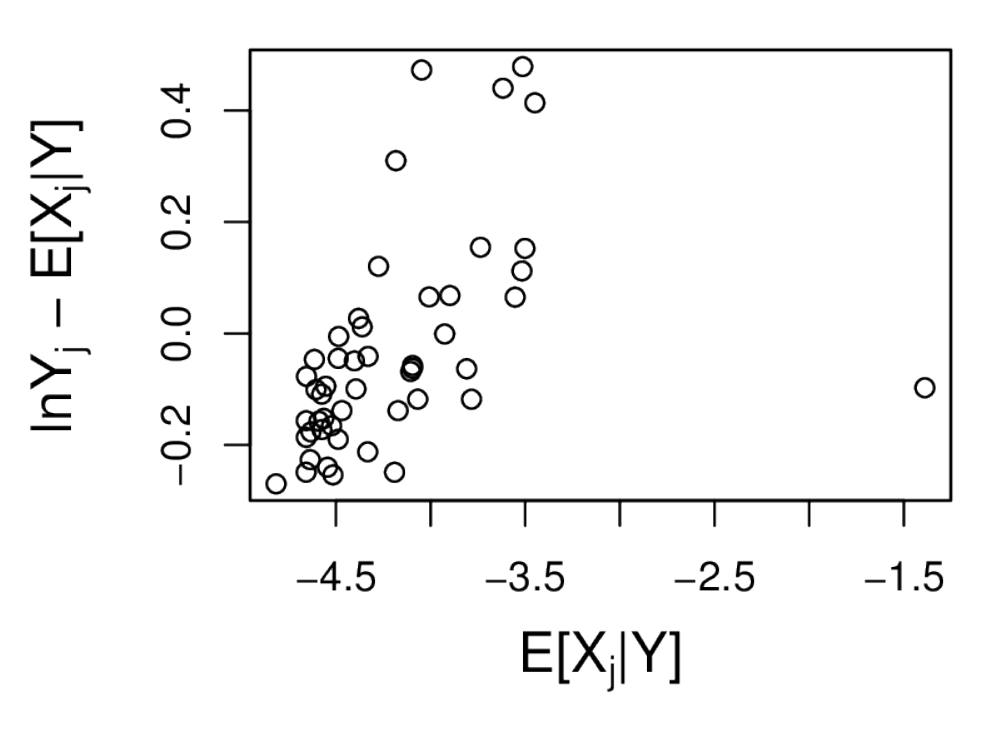}
\includegraphics[width=1.\textwidth]{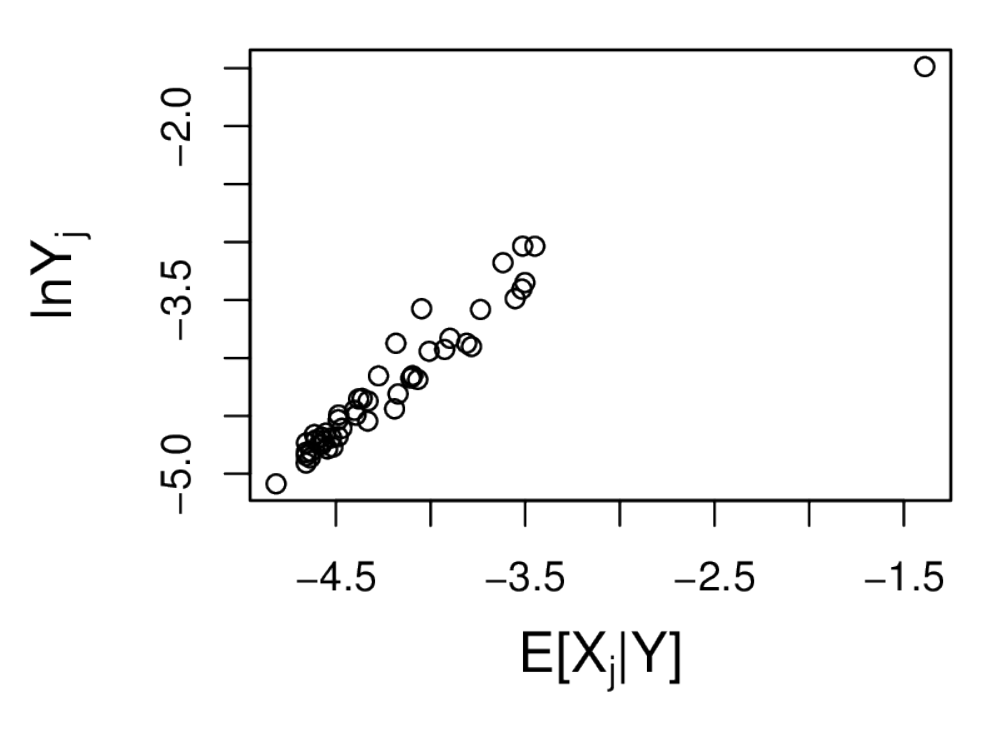}
	\end{minipage}
	\caption{Visualisation of the posterior latent variables $X$ for the low-noise regime ($\gamma=10^4$). Left:  The outer and inner rings show the posterior mean plus or minus three standard deviations, respectively.  Top-right: Expected attractiveness against the expected residual.  Bottom-right: Expected attractiveness against observed value.}
\label{low-lat}
\end{figure}

 Similar plots are presented in Figures.~\ref{low-density}-\ref{low-lat} for the low-noise regime, and the posterior-marginals of $\alpha$ and $\beta$ give a mean of $1.17 \pm 0.01$ and $0.26 \pm 0.01$, respectively.  The inferred values for the low-noise regime are in line with Figure~\ref{opt_results}.   Both sets of posterior summaries suggest that attractiveness and inconvenience effects are present in the data, though the inferred $\alpha$ values are considerably higher for the low-noise model. The plots of the expected residuals and observation data suggest that the model also provides a reasonable fit to the data, however, there is more dispersion in the plotted quantities and possibly a degree of heteroscedasticity due to the less flexible model. The model with more noise favours the simpler explanation that most variation is due to stochastic growth, whereas the low-noise model is more constrained.  There is noticeably more uncertainty in the latent variables and the model parameters for the high-noise regime, as there are more possible explanations for the observation data.   The uncertainty in the $\alpha$ and $\beta$ estimates is so great for the high-noise regime that limited insights are gained for the purposes of model calibration.  As a result, we conclude that strong assumptions are required in the prior modelling in order to be able to exploit known structure in the data generating process.  The required assumptions can be made, for example, through the prior modelling of $\alpha$ and $\beta$ or by specifying a high value of $\gamma$.  On the other hand, the low-noise regime results in very confident posteriors.  Although the resulting inferences are consistent with previous studies, care must be taken to avoid being overconfident in a particular model by not adequately accounting for uncertainty in the modelling process.

\section{Discussion}
\label{S:diss}
We have developed a novel stochastic model to simulate realistic configurations of urban and regional structure.  Our model is a substantial improvement on existing deterministic models as it fully addresses the uncertainties arising in the modelling process.  Unlike existing time-stepping schemes, our model can be used to simulate realistic configurations of urban structure using MCMC methods without recourse to numerical error.  We have demonstrated that our model can be used to infer the components of a utility function from observed structure, thereby providing an alternative to the existing discrete choice models.  The key advantage is that we avoid the need to collect vast amounts of flow data.  Whilst we have presented our methodology in the context of consumer-led behaviour, our approach is applicable to other urban and regional settings such as archaeology, logistics, health care and crime to suggest a few. 

Our work has led to specific areas for further research.  We are actively investigating the deployment of our methodology to large scale urban systems, for which there are substantial computational challenges to overcome.  The cost of a potential or gradient evaluation is $O(NM)$, however, increasing $M$ means that the $z(\theta)$ estimates are more challenging to obtain owing to the curse of dimensionality.  It is of interest to develop more tractable methods, for example, optimisation-based, so that inference can be performed for international models on a practical time scale.  We have presented an aggregate model that can be refined to better represent domain-specific characteristics as discussed in~\cite{dearden2015explorations}.  It remains to use the proposed methodology as part of a more realistic study with wider objectives.  Lastly, we emphasise that our methodology is only applicable to cross sectional data.  In practice, many applications of interest require processing time-series data that is highly correlated over time.  In this setting, we would need to solve the filtering or smoothing problem for~\eqref{stochastic_harris_wilson2}, and in doing so would also need to account for general trends and seasonality effects that are exogenous to our model.   Our work continues to be part of ongoing efforts to draw insights from data by making use of the known mathematical structure~\cite{apte2008data}.  

\appendix 
\numberwithin{equation}{section}

\section{Assumptions for the potential function}
\label{Appendix:Assumptions}

We make the following assumptions for the potential function $V(x)$ with reference to the overdamped Langevin diffusion described by~\eqref{lang} and Boltzmann-Gibbs measure defined by~\eqref{BoltzmannGibbs}:
\begin{enumerate}
	\item $V(x)$ is $C^2$ and is confining in that $\lim_{| x | \to + \infty} V(x) = + \infty$ and
	\begin{equation}
	e^{-\gamma V(x)} \in L^1(\mathbb{R}^M), \quad \forall \, \gamma > 0.
	\end{equation}
	\item $V(x)$ satisfies the following inequality for some $0 < d < 1$
	\begin{equation}
	\liminf_{|x| \to \infty} \bigg\{ (1-d) \big|  \nabla V(x) \big|^2 - \gamma^{-1} \Delta V(x) \bigg\} > 0.
	\end{equation}
\end{enumerate}
Assumption (i) is necessary to ensure that the Boltzmann-Gibbs measure is well-defined.  Assumptions (i) and (ii) are sufficient to show that the distribution of $X(t)$, described by~\eqref{lang}, converges exponentially fast to its Boltzmann-Gibbs measure. The reader is referred to~\cite{roberts1996exponential} for further details.

The integrability condition in assumption (i) is satisfied by Lemma 3.14 in~\cite{menz2014poincare}.  To show that assumption (ii) holds, we define
\begin{equation}
\Lambda_{ij} = \frac{ \exp \big(\alpha x_j - \beta c_{ij} \big) }{\sum_{k=1}^M \exp \big(\alpha x_k - \beta c_{ik} \big)},
\end{equation}
then
\begin{equation}
\begin{aligned}
\big| \nabla V(x) \big|^2 & = 
\sum_{j=1}^M \bigg|    \sum_{i=1}^N O_i \Lambda_{ij} - \kappa \exp (x_j ) +  \delta \bigg|^2,
\end{aligned}
\end{equation}
and
\begin{equation}
\label{hessian}
\begin{aligned}
\Delta V(x) = 
\sum_{j=1}^M \Bigg\{ \kappa \exp (x_j ) -  \alpha \sum_{i=1}^N O_i  \Lambda_{ij} \big( 1 - \Lambda_{ij} \big)  \Bigg\}.
\end{aligned}
\end{equation}
Then for $0 < d < 1$ we have
\begin{equation}
\lim_{x_j \to +\infty} (1 -d) \big| \nabla V(x) \big|^2 - \gamma^{-1}\Delta V(x) = +\infty,
\end{equation}
and
\begin{equation}
\lim_{x_j \to -\infty}  (1 - d) \big| \nabla V(x) \big|^2 - \gamma^{-1} \Delta V(x) \geq \delta^2 > 0,
\end{equation}
as claimed.

\section{Utility function for a singly-constrained potential}
\label{Appendix:Utility}
In this appendix, we present an alternative argument to obtain the utility potential  $V_{\text{Utility}}$ as defined by~\eqref{path}.  We rewrite the destination quantities in~\eqref{d_eq} as a gradient flow
\begin{equation}
\label{targ-goal}
\begin{aligned}
D_j = - \frac{\partial V_{\text{Utility}}(x)}{\partial x_j},
\end{aligned}
\end{equation}
and look towards specifications of $ V_{\text{Utility}}$ that satisfy the constraint in~\eqref{o_eq}.  The constraint is satisfied whenever the flows leaving the origin zones are convex sums in that
\begin{equation}
\label{rev1}
T_{ij}(x) =  O_i v_{ij}(x), \qquad \sum_{j=1}^M v_{ij} = 1, \qquad v_{ij} \geq 0.
\end{equation}
Instead we can express~\eqref{rev1} in terms of utility functions $U_{ij}$ and some positive function $\varphi$ so that
\begin{equation}
\label{dem2}
T_{ij} = O_i \frac{\varphi \big(U_{ij} \big)}{\sum_{k=1}^{M}\varphi \big(U_{ik} \big)}.
\end{equation}
By inspection, we look for a potential function of the form
\begin{equation}
V_{\text{Utility}} = - {\sum_{i=1}^N O_i  \Bigg\{ f_i  \ln \sum_{j=1}^M \varphi \big( U_{ij} \big)} \Bigg\},
\end{equation}
for some functions $f_i$.  Then by taking the gradient and substituting into~\eqref{targ-goal}, we obtain the requirements
\begin{equation}
\begin{aligned}
\label{req}
\frac{\varphi \big(U_{ij} \big)}{\sum_{k=1}^{M}\varphi \big(U_{ik} \big)} = & \, \frac{d f_i}{dx_j} \ln \sum_{j=1}^M \varphi \big( U_{ij} \big)  +  f_i \frac{d \varphi\big(U_{ij} \big)}{d x_j} \bigg( \sum_{k=1}^{M}\varphi \big(U_{ik} \big) \bigg)^{-1},
\end{aligned}
\end{equation}
for $i = 1, \dots, N$.  The requirements are satisfied for $\varphi(\cdot) = \exp(\cdot)$ when each utility function is a linear with respect to the attractiveness of the destination zone in question,
\begin{equation}
\label{utility}
U_{ij} = \alpha_i x_j + \beta_{ij},
\end{equation} 
provided that each $\alpha_i \neq 0$ and that $f_i = \alpha_i^{-1}$.  The resulting potential function is
\begin{equation}
V_{\text{Utility}}(x) = -{ \sum_{i=1}^N \alpha_i^{-1 } {O_i}  \ln \sum_{j=1}^M \exp \big( U_{ij} \big)},
\end{equation}
which is slightly more general than the potential considered in~\eqref{path}.

\section{Markov chain Monte Carlo for doubly intractable distributions}
\label{Appendix:MCMC}
The idea behind MCMC is to construct an ergodic Markov chain whose time-average can be used to estimate integrals of interest~\cite{liu2008monte, robert2004monte}.  We use a block Metropolis-within-Gibbs scheme and alternate between $\Theta$-updates and $X$-updates.  The following steps can be repeated in succession to obtain a Markov-chain chain that is $\pi(x, \theta | y)$-invariant:

\paragraph{Latent variable update.}  Hamiltonian Monte Carlo can be used for the $X$-updates to suppress random walk behaviour~\cite{neal2011mcmc}.  We propose momentum variables $P \sim N(0, I)$ and update $(X, P)$ by simulating Hamiltonian dynamics with a volume-preserving integrator to obtain $(X', P')$.  We accept/reject according to the Metropolis-Hastings acceptance probability
\begin{equation}
a_X(x', p' | x, p) = \min \bigg\{ 1, \frac{ \pi(y | x', \theta) \, \exp \big(- \gamma V_\theta (x')  - \frac{1}{2} | p' |^2 \big) }{ \pi(y | x, \theta) \,  \exp \big(- \gamma V_\theta (x) - \frac{1}{2} | p|^2 \big)  } \bigg\},
\end{equation}
to correct for numerical error from numerically simulating Hamiltonian dynamics.

\paragraph{Model parameter update.}  Random walk Metropolis with reflective boundaries can be used for the $\Theta$-updates.   The Metropolis-Hastings acceptance probability is given by
\begin{equation}
a_\Theta(\theta' | \theta) = \min \bigg\{ 1, \frac{ \pi(y | x', \theta' ) z(\theta) \exp \big( -\gamma V_{\theta'}(x) \big) \pi(\theta') }{ \pi(y | x, \theta) z(\theta') \exp \big( -\gamma V_{\theta}(x) \big) \pi(\theta) }\bigg\}.
\end{equation}
The key challenge arises from proposing new $\Theta$-values, in which case the acceptance probability contains an intractable ratio $z(\theta)/z(\theta').$  We can either proceed with a deterministic estimate of $z(\theta)$, at the expense of a bias,  or we can obtain a consistent estimator of~\eqref{obj} with pseudo-marginal MCMC~\cite{andrieu2009pseudo} provided that unbiased and reasonably precise estimates of $1/z(\theta)$ are available.

\subsection{Unbiased estimates of the reciprocal of the normalizing constant}
An unbiased estimate  of $z(\theta)$ is given by averaging over a batch of importance weights.  The importance weights are evaluations of
\begin{equation}
\label{impw}
w(x) = \exp \big( -\gamma V_\theta(x) \big) / q(x),
\end{equation}
at locations drawn from a proposal distribution with density $q(x)$.  By Jensen's inequality, the reciprocal of an importance sampling estimate of $z(\theta)$ is a biased estimate of $1/z(\theta)$.  Instead, unbiased estimates of $1/z(\theta)$ can be obtained by randomly truncating an infinite series:  for a sequence $\{ \mathcal{V}_i \}$ satisfying $\lim_{i \to + \infty} \mathbb{E} \big[ \mathcal{V}_{i} \big] = 1/z(\theta)$, and a random stopping time $T$, the estimator
\begin{equation}
\label{roulette3}
S = \mathcal{V}_0 + \sum_{i=1}^{K} \frac{ \mathcal{V}_i- \mathcal{V}_{i-1}}{\text{Pr}\big(T \geq i \big)},
\end{equation}
gives an unbiased estimate of $1/z(\theta)$~\cite{lyne2015russian, mcleish2010general, glynn2014exact, wei2016markov}.  We follow~\cite{murray2006} and use the increasing averages estimator
\begin{equation}
\mathcal{V}_i = \frac{i+1}{\sum_{k=0}^i w \big(\hat{X}^{(k)} \big)}, \qquad \hat{X}^{(k)} \sim q,
\end{equation}
with reference to the importance sampling weights described by~\eqref{impw}.  The unbiased estimates of $1/z(\theta)$ can have high variance when the importance weights are highly variable.  Fortunately, importance sampling may be carried out on an augmented state space, for example, using annealed importance sampling (AIS)~\cite{neal2001annealed}.

\subsection{Pseudo-marginal Markov chain Monte Carlo}
The unbiased estimators of $1/z(\theta)$ given by~\eqref{roulette3} may be negative estimate, which prohibits the use of pseudo-marginal MCMC.  Fortunately, the so called `sign problem' has been addressed in~\cite{lyne2015russian}.  We can use the following importance sampling style of estimator that gives a consistent estimator of~\eqref{obj} in that
\begin{equation}
\mathbb{E}_{x, \theta | y}[g] =  \lim_{n \to + \infty} \frac{\sum_{i=1}^n \Omega^{(i)} g\big(X^{(i)}, \Theta^{(i)}\big)}{\sum_{k=1}^n \Omega^{(k)} },
\end{equation}
where $\big\{ X^{(i)}, \Theta^{(i)}, \Omega^{(i)} \big\}_{i=1}^n$  is a Markov-chain obtained using the Metropolis-within-Gibbs scheme described at the start of this section but with the following acceptance probability for $\Theta$-updates
\begin{equation}
a_\Theta(\theta' | \theta) = \min \bigg\{ 1, \frac{ \pi(y | x', \theta' ) |S'| \exp \big( -\gamma V_{\theta'}(x) \big) \pi(\theta') }{ \pi(y | x, \theta) |S|  \exp \big( -\gamma V_{\theta}(x) \big) \pi(\theta) }\bigg\}.
\end{equation}
and $\Omega^{(i)} = \text{sgn} \big( S^{(i)} \big)$.  It is necessary to cache the value of $S^{(i)}$ at each iteration as part of the pseudo-marginal MCMC scheme.

\section{Table of key parameters and variables}
\label{Appendix:Params}
For convenience, we provide a table of the key parameters and variables with brief explanations below:
\begin{center}
	\begin{tabular}{ |c|l|c| } 
		\hline
		Parameter & Explanation & Reference \\
		\hline
		$\alpha$ & Attractiveness scaling parameter. &\eqref{flow} \\ 
		$\beta$ & Cost scaling parameter. & \eqref{flow} \\ 
		$\delta$ & Additional parameter. & \eqref{pot_sum} \\ 
		$\epsilon$ & Responsiveness parameter.&\eqref{HW} \\ 
		$\gamma$ & Inverse-temperature.&\eqref{lang} \\ 
		$\kappa$ & Cost per unit size.&\eqref{HW} \\ 
		$\lambda$ & Standard deviation of observation noise.&\eqref{random_like} \\ 
		$\sigma$ & Noise parameter, equal to $\sqrt{2\gamma^{-1}}$.& \eqref{stochastic_harris_wilson2}\\ 
		$O_i$ & Origin quantity for origin zone $i$.&\eqref{o_eq} \\ 
		$D_j$ & Destination quantity for destination zone $j$.& \eqref{d_eq}\\ 
		$T_{ij}$ & Flow from origin zone $i$ to destination zone $j$.& \eqref{o_eq}\\ 
		$U_{ij}$ & Utility function for a flow from origin zone $i$ to destination zone $j$.&\eqref{utility_func} \\ 
		$c_{ij}$ & Cost of a flow from origin zone $i$ to destination zone $j$.&\eqref{flow} \\ 
		$W_j$ & Size of destination zone $j$.& \eqref{HW}\\ 
		$X_j$ & Attractiveness of destination zone $j$, given by $\ln W_j$.&\eqref{lang} \\ 
		$Y_j$ & Observed size of destination zone $j$.& \eqref{random_like} \\ 
		$N$ & Number of origin zones.& \eqref{o_eq} \\ 
		$M$ & Number of destination zones.& \eqref{d_eq}\\ 
		\hline
	\end{tabular}
\end{center}

\enlargethispage{20pt}

\section*{Acknowledgements} L.E. was supported by EPSRC [EP/P020720/1]. MG was supported by EPSRC [EP/J016934/3, EP/K034154/1, EP/P020720/1, EP/R018413/1], an EPSRC Established Career Fellowship and The Alan Turing Institute - Lloyd's Register Foundation Programme on Data-Centric Engineering. G.P. was supported by EPSRC [EP/P031587, EP/L020564, EP/L024926, EP/L025159]. A.W. was supported by EPSRC [EP/M023583/1]. This work was supported by The Alan Turing Institute under the EPSRC grant EP/N510129/1.

\vskip2pc

\bibliographystyle{unsrt}
\bibliography{refs}

\end{document}